\documentclass[twocolumn]{emulateapj}
\shorttitle{SB2s}
\shortauthors{Fernandez et al.}

\begin{document}


\title{IN-SYNC VI. Identification and Radial Velocity Extraction for 100+ Double-Lined Spectroscopic Binaries in the APOGEE/IN-SYNC Fields}
\author{M.A. Fernandez\altaffilmark{1,a}, Kevin R. Covey\altaffilmark{1,b}, Nathan De Lee\altaffilmark{2,3}, S. Drew Chojnowski\altaffilmark{4,5},
David Nidever\altaffilmark{6},
Richard Ballantyne\altaffilmark{1},
Michiel Cottaar\altaffilmark{7},
Nicola Da Rio\altaffilmark{8},
Jonathan B. Foster\altaffilmark{9,10},
Steven R. Majewski\altaffilmark{5},
Michael R. Meyer\altaffilmark{11},
A.M. Reyna\altaffilmark{1},
G.W. Roberts\altaffilmark{1}, 
Jacob Skinner\altaffilmark{1},
Keivan Stassun\altaffilmark{3},
Jonathan C. Tan\altaffilmark{8},
Nicholas Troup\altaffilmark{5},
Gail Zasowski\altaffilmark{12}}
\altaffiltext{1}{Western Washington University, 516 High St., Bellingham, WA 98225-9164 $^a$current email: mfern027@ucr.edu, $^b$ kevin.covey@wwu.edu} 
\altaffiltext{2}{Northern Kentucky University, Nunn Drive, Highland Heights, KY 41099}
\altaffiltext{3}{Department of Physics \& Astronomy, Vanderbilt University, Nashville, TN 37235}
\altaffiltext{4}{Department of Astronomy, New Mexico State University, 1780 E University Ave, Las Cruces, NM 88003}
\altaffiltext{5}{Department of Astronomy, University of Virginia, Charlottesville, VA 22904}
\altaffiltext{6}{National Optical Astronomy Observatory, 950 North Cherry Ave, Tucson, AZ 85719}
\altaffiltext{7}{Department of Clinical Neurosciences, University of Oxford, Oxford, United Kingdom}
\altaffiltext{8}{Departments of Astronomy and Physics, University of Florida, Gainesville, FL 32611, USA}
\altaffiltext{9}{Yale Center for Astronomy and Astrophysics, Yale University New Haven, CT 06520, USA}
\altaffiltext{10}{Legendary Entertainment Applied Analytics 535 Boylston St. Suite 401 Boston, MA 02116}
\altaffiltext{11}{Department of Astronomy, University of Michigan, 1085 S. University, Ann Arbor, MI 48109}
\altaffiltext{12}{Department of Physics \& Astronomy, Johns Hopkins University, Baltimore, MD 21218, USA}

\begin{abstract}
We present radial velocity measurements for 70 high confidence, and 34 potential binary systems in fields containing the Perseus Molecular Cloud, Pleiades, NGC 2264, and the Orion A star forming region. 18 of these systems have been previously identified as binaries in the literature. Candidate double-lined spectroscopic binaries (SB2s) are identified by analyzing the cross-correlation functions (CCFs) computed during the reduction of each APOGEE spectrum. We identify sources whose CCFs are well fit as the sum of two Lorentzians as likely binaries, and provide an initial characterization of the system based on the radial velocities indicated by that dual fit. For systems observed over several epochs, we present mass ratios and systemic velocities; for two systems with observations on eight or more epochs, and which meet our criteria for robust orbital coverage, we derive initial orbital parameters. The distribution of mass ratios for multi-epoch sources in our sample peaks at q=1, but with a significant tail toward lower q values. Tables reporting radial velocities, systemic velocities, and mass ratios are provided online. We discuss future improvements to the radial velocity extraction method we employ, as well as limitations imposed by the number of epochs currently available in the APOGEE database. The Appendix contains brief notes from the literature on each system in the sample, and more extensive notes for select sources of interest.
\end{abstract}


\section{Introduction} \label{introduction}

The frequency and architecture of stellar binaries are a fundamental outcome of the star formation process. Computational models of molecular cores and star forming regions make quantitative predictions for the multiplicity fraction and orbital architectures of pre-main sequence binaries \citep[e.g.][]{Bate2012, Parker2014, Lomax2015}.  Empirical measurements of binary properties provide a key means of testing such models, but a variety of observational techniques are required to identify and characterize multiple systems across the full range of primary masses, orbital separations, mass ratios and ages/evolutionary states. Recent efforts include direct imaging surveys, both seeing-limited \citep{Kraus2007, Connelley2008} and with adaptive optics \citep{Duchene1999, Correia2006, Lafreniere2008, Kraus2009, Vogt2012, Ward-Duong2015}, as well as imaging enhanced with speckle \citep[e.g.][]{Ghez1993, Ghez1997, Ratzka2005}, aperture masking \citep{Kraus2008, Kraus2011, Cheetham2015} and multi-aperture interferometric \citep{Guenther2007} techniques. 

Spectroscopic surveys are required to identify systems with the smallest separations. Double-lined spectroscopic binaries (SB2s) allow us to measure the motions of each component separately, providing a robust measurement of the binary systemic velocity and mass ratio; for systems viewed edge-on, absolute masses for each component can also be derived.  Previous surveys to identify pre-main sequence spectroscopic binaries have been conducted at optical \citep{Melo2003, Nguyen2012, Kounkel2016, Kohn2016} and infrared \citep{Prato2007, VianaAlmeida2012} wavelengths. Infrared surveys are typically less efficient than optical surveys, due to higher overheads associated with telluric corrections and flux calibration, but are sensitive to cooler companions, and less evolved, optically fainter systems. 

The spectra collected by the INfrared Spectra of Young Nebulous Clusters (IN-SYNC) SDSS-III ancillary science project provide a rich resource for identifying and characterizing pre-main sequence binaries in nearby star forming regions. The high-resolution near-infrared spectra obtained with the Apache Point Observatory Galactic Evolution Experiment (APOGEE) multi-object spectrograph enable efficient measurement of precise ($\sigma_{RV} <$ 1 km s$^{-1}$) radial velocities (RVs) for thousands of pre-main sequence stars and protostars over multiple epochs. Analysis of the RVs extracted from these spectra have already been used to diagnose the dynamics of several nearby star-forming regions, including IC 348 \citep{Cottaar2014, Cottaar2015}, NGC 1333 \citep{Foster2015}, and the Orion A filament \citep{DaRio2016,DaRio2017}.

In this work, we analyze the cross-correlation functions (CCFs) measured from each individual APOGEE spectrum to identify and characterize SB2s located within the IN-SYNC fields. Specifically, we measure RVs from each APOGEE spectrum for each component in a system, which we use to characterize the system's properties. For two systems in our sample we determine all seven orbital parameters; for the majority of the multi-epoch systems in our sample, which have only 2-7 epochs of observations, we only attempt to measure the system's mass ratio and systemic velocity.

In Section \ref{observations} we briefly overview the target selection, spectroscopic observations, and APOGEE pipeline, which produces the CCFs used throughout. In Section \ref{rvs} we detail the procedure with which we extract radial velocities (\ref{extraction}), discuss the process of identifying SB2s in our sample (\ref{selection}), and discuss the RV residuals, a pseudo-measure of the uncertainty (\ref{residuals}). We present our results, including radial velocities, systemic velocities, mass-ratios, and orbit fits in Section \ref{results}. In Section \ref{conclusion} we look toward future work, laying out improvements, as well as discussing limitations imposed on this procedure by the number of epochs for sources in the sample. In the Appendix we present literature and cluster membership notes on each system.

\section{Data} \label{observations}

\subsection{APOGEE/IN-SYNC Observations} \label{spectra}

We analyze high-resolution (R$\sim$ 22,500) near-infrared ($1.51$-$1.70$ $\mu m$) spectra taken with the 300-fiber APOGEE spectrograph \citep{Wilson2012} on the Sloan 2.5 meter telescope \citep{Gunn2006} at Apache Point Observatory.  During SDSS-III, the APOGEE survey obtained nearly 620,000 spectra of more than 150,000 stars throughout the Milky Way, which were released in the SDSS twelfth data release \citep[DR12; ][]{Alam2015}. The vast majority of APOGEE targets are red giants \citep{Zasowski2013}, collected in service of the survey's primary science goal, to dissect the structure, dynamics, chemical evolution and star formation history of the Milky Way \citep{Majewski2015}.  As \citet{Alam2015} describe, several `ancillary science' programs were approved to utilize the unique capabilities of the APOGEE spectrograph and data analysis infrastructure to target complimentary science goals.  Our analysis focuses on identifying SB2s within fields observed as part of the IN-SYNC ancillary science program, which targeted regions of recent and current star formation activity.  

\subsection{The IN-SYNC Sample}
The primary regions targeted by the IN-SYNC program are IC 348 \citep{Cottaar2014, Cottaar2015} and NGC 1333 \citep{Foster2015} in the Perseus Molecular Cloud (PMC), the Orion A molecular cloud \citep{DaRio2016} and NGC 2264. We summarize here the most pertinent details of the target selection and observations in the fields containing these regions; for more details we refer the reader to the full descriptions given in each of the papers cited above.  The highest priority targets in each region were selected from existing catalogs of young stellar objects (YSOs) identified by photometric or spectroscopic signatures of youth, such as infrared excess, emission line activity, and/or low surface gravity.  If high-priority targets did not require the full complement of 230 science fibers allocated to each APOGEE field, candidate members selected via color-magnitude or proper motion cuts were allocated fibers at lower priority. In total, nearly 3500 pre-main sequence stars were observed as part of the IN-SYNC program: ∼2700 in the Orion A cloud, ∼380 in IC 348, and ∼110-120 in NGC 1333 and NGC 2264.  For this paper, however, we have analyzed all sources in the fields containing the IN-SYNC targets, regardless of how they were originally targeted, as some may provide serendipitous detections of cluster members. Including all targets in each field provides a final sample of 4556 unique sources: 2771 in Orion, 1309 in Perseus (IC 348 and NGC 1333), 100 in Pleiades, and 376 in NGC 2264: for the remainder of this paper, we refer to these objects as the Complete IN-SYNC Sample.

\begin{figure}
\centering
\includegraphics[width = 0.8\hsize]{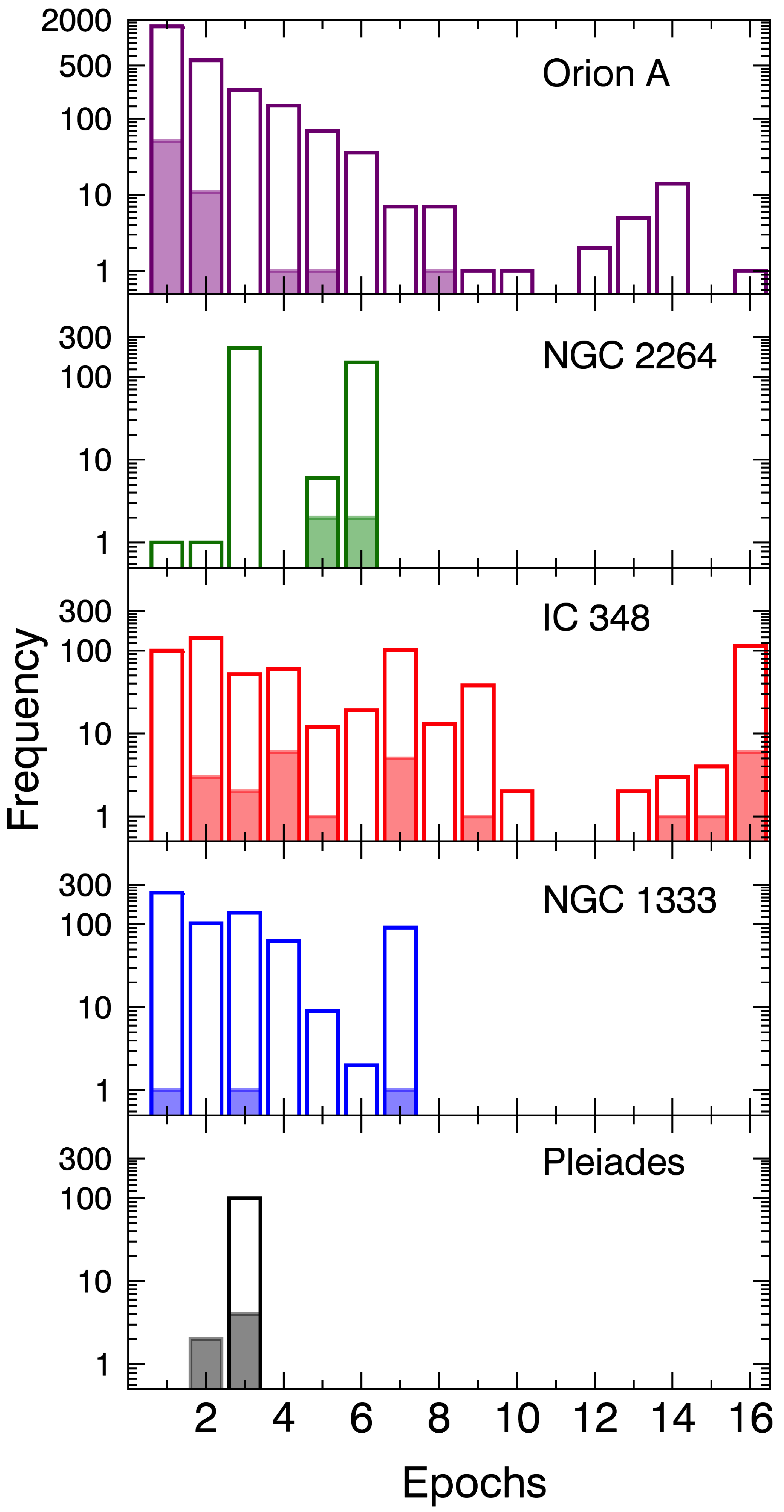}
\caption{Distribution of the number of epochs observed for binaries in each of the IN-SYNC fields for the Complete IN-SYNC sample (unshaded) and the IN-SYNC SB2 sample (shaded). Some epochs were excluded due to low S/N or poorly fit CCFs (Section \ref{extraction}, \ref{selection}), so some sources have fewer epochs in the IN-SYNC SB2 sample than in the Complete IN-SYNC sample. The Complete IN-SYNC sample also contained two 17 epoch and three 18 epoch sources in Orion. Both of the systems for which we were able to fit orbits (Section \ref{orbits}) are 16 epoch sources located in the IC 348 field.}
\label{fig:cluster_epochs}
\end{figure}

As seen in Figure \ref{fig:cluster_epochs}, the observing strategy and schedule achieved in different IN-SYNC target regions resulted in distinct cadences for each of the binaries identified in our sample. Fibers cannot be placed on targets separated by less than 72\arcsec on a single APOGEE plate, such that repeat observations of a given APOGEE field are required to observe targets with nearby neighbors.  Fields in Orion, which were limited to a narrow observing window, emphasized using repeat visits to a given field to alleviate crowding and expand the total sample size, such that a typical Orion target was only observed for one epoch. The IC 348 field, by contrast, was observed over a longer window, with visits dedicated to obtaining multiple epochs for sources at intermediate and large cluster radii, to enable the detection and characterization of binaries in the cluster and surrounding field populations.  Sources in the IC 348 field were therefore typically observed at least 3-4 times, with a substantial number having as many as 16 observations.  

\subsection{APOGEE Pipeline} \label{apogee}

The APOGEE pipeline produces reduced spectra for each observation of a target, as well as a combined spectrum co-adding all observations to maximize the resultant signal-to-noise.  We focus our analysis on the individual spectra of each source, typically referred to as `visit spectra' in the APOGEE documentation, as these time-resolved observations are of greater use for identifying and characterizing SB2 systems.  

\begin{figure*}
\centering
\includegraphics[width = 0.75\hsize]{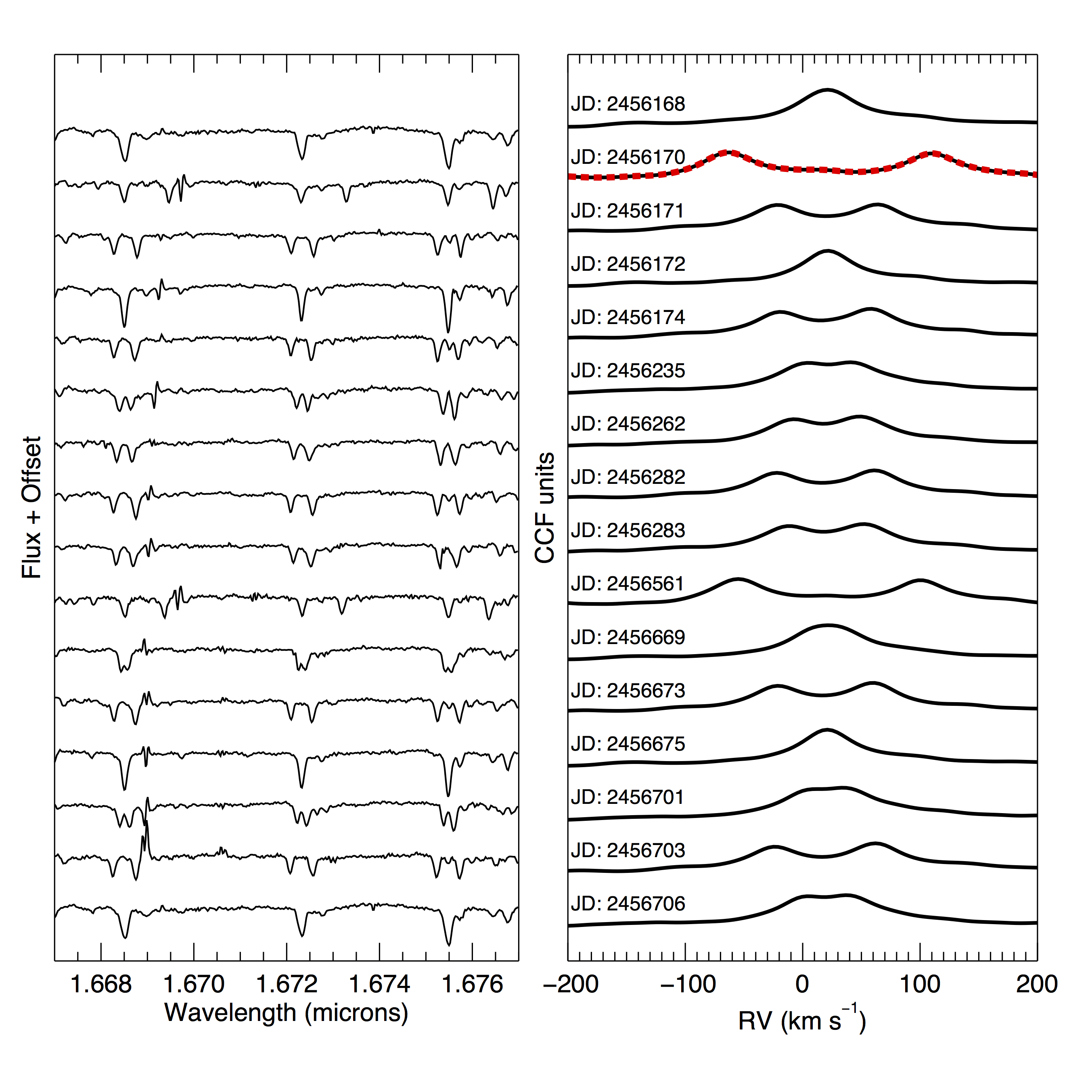}
\caption{Left panel: Small section of APOGEE spectra for 2MJ03434101+3237320, a newly identified SB2. Right panel: Stacked CCFs, produced by the APOGEE pipeline, for the same system. Julian date is indicated for each epoch. The dashed CCF highlighted in red is the widest separated epoch, the use of which is described in Section \ref{extraction}. Three of these CCFs, with three differing morphologies and RV fits are shown in Figure~\ref{fig:fits}.}
\label{fig:stackCCF}
\end{figure*}

The APOGEE pipeline calculates CCFs for each visit spectrum (see Figure \ref{fig:stackCCF} for sample spectra and associated CCFs for a high-visit IN-SYNC target in IC 348). The spectrum is cross-correlated against the best fit synthetic spectrum from the APOGEE RV mini-grid, a set of models sparsely spanning effective temperature, metallicity, and log(g) space. The resulting CCF is computed for 401 lags, with a velocity resolution per lag step of 4.14 km s$^{-1}$; as \citet{Nidever2015} and \citet{Cottaar2014} demonstrate, however, the stability of the APOGEE spectrograph enables the measurement of RVs at the 0.1-0.2 km s$^{-1}$ level by centroiding each CCF peak.

The APOGEE reductions of each visit spectrum provide the lag-to-velocity conversion factor, the heliocentric correction for that epoch, and the radial velocity estimate determined by the APOGEE pipeline.  We do not utilize the radial velocity measurements produced by the APOGEE pipeline, as those measurements are derived under the assumption that the source in question is a single star. Instead, we perform a new analysis of the APOGEE-produced cross-correlation functions to identify likely SB2s and determine each component's radial velocity.

\section{Radial Velocities} \label{rvs}

Accurately extracting multiple RVs from thousands of APOGEE spectra requires that we reliably fit multi-component CCFs en masse. The process of fitting CCFs is the topic of Section \ref{extraction}, while Section \ref{selection} details how we use these measurements to identify SB2s in an efficient, semi-automated procedure. Section \ref{residuals} presents our measure of uncertainty for the extracted RVs.

\subsection{RV Extraction} \label{extraction}

To extract radial velocities we begin by converting the APOGEE pipeline CCFs from lag-space to velocity-space, using:
\begin{equation}
v_{radial} = (10^{\alpha \beta} - 1) c + v_{helio},
\end{equation}

\noindent where $c$ is the speed of light in vacuum, $\beta$ gives the index for the array of 401 CCF lag steps, $v_{helio}$ is that star's heliocentric correction at the epoch of observation, and $\alpha$ is the exponential lag-to-velocity conversion factor ($6 \times 10^{-6}$ or $4.14$ km s$^{-1}$ per step) associated with the APOGEE-produced CCFs.

Our dual RV-extraction algorithm first attempts to identify two distinct velocity components in the CCF measured from each visit spectrum. The maximum of each CCF is identified as the primary peak, and the portion of the CCF bounded by the local minima on either side of the primary peak is removed. Once the primary peak has been subtracted from the CCF, the maximum of the remaining portion of the CCF is selected as a candidate secondary peak, and the velocity separation between the primary and putative secondary peak is calculated to characterize the velocity separation at that epoch. Comparing the velocity separations measured between the primary and secondary peaks for all observations of a given source, the epoch with the largest peak separation can be identified; in the discussion that follows, we identify this epoch as the `widest separated CCF'.  This procedure forces a dual-peak fit to all CCFs, even sources which have only one astrophysically significant peak. The parameter cuts detailed in Section \ref{selection}, however, eliminate most sources with genuinely singly-peaked CCFs from our final sample of candidate SB2s. 

\begin{figure}
\centering
\includegraphics[width = 0.8\hsize]{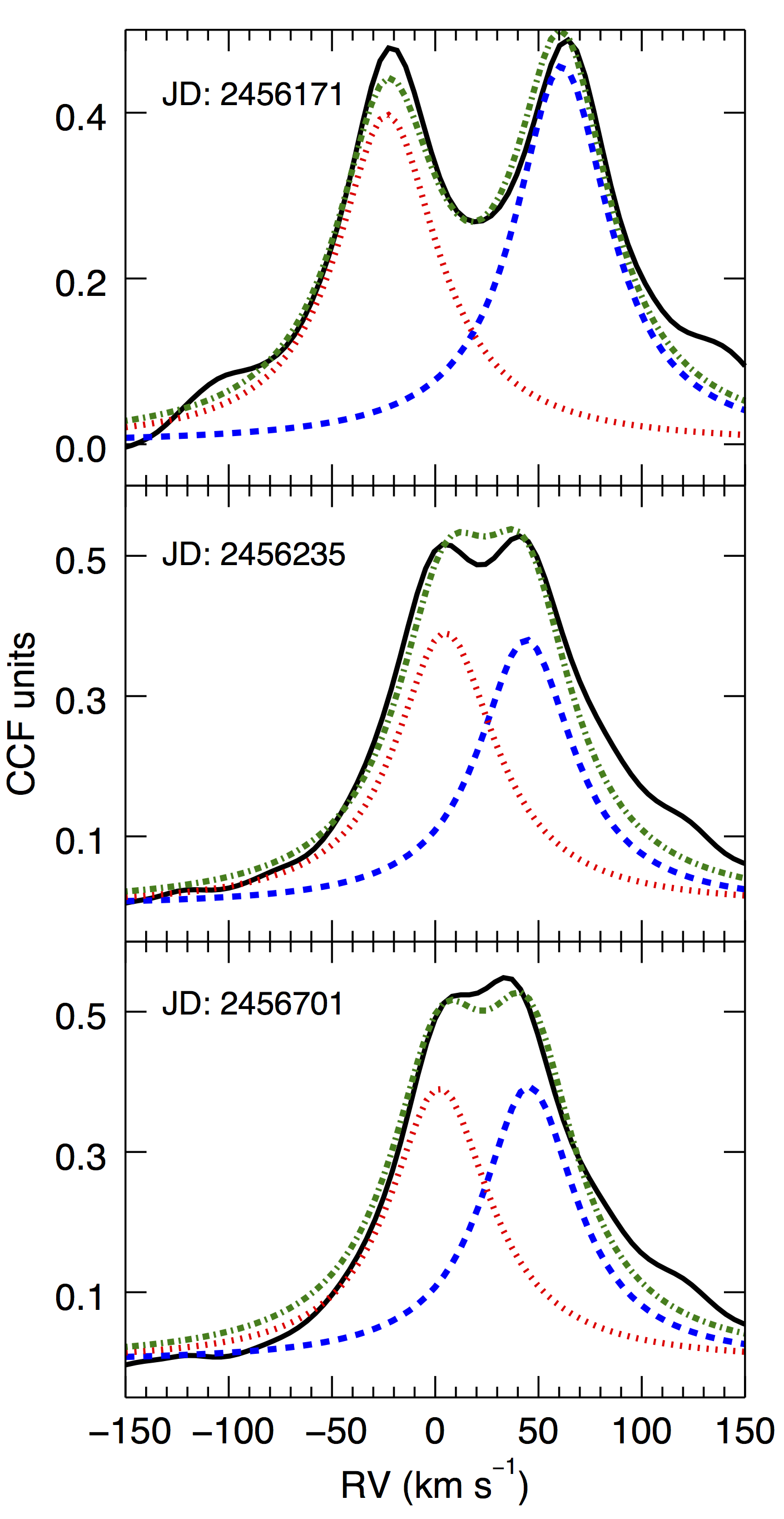}
\caption{Example cross-correlation functions (black) with individual (blue and red) and dual (green) Lorentzian fits  from the system 2MJ03434101+3237320. The fit peak locations correspond to the extracted radial velocities. From top to bottom panel we have an example of: separated peaks (reliable indicator of binarity), merged structures (likely indicator of binarity), and shoulder behavior (possible indicator of binarity).}
\label{fig:fits}
\end{figure}

Once a source's widest separated CCF has been identified, Lorentzians are fit to all epochs of the source's CCF in three steps. First, the components of the widest separated CCF are fit separately to establish an initial set of parameters (i.e. central velocity, peak height, and width parameter) to describe each peak. Next, using the best-fit parameters for each peak as a starting point, a second fit to the widest separated CCF is then performed, with both peaks fit simultaneously. In the third and final step, the best-fit parameters from the dual fit to the widest separated CCF are used to initiate dual component fits to all of a source's CCFs. Detailed descriptions of each step follow:

\begin{enumerate}
\item \textbf{Individual fits to primary and secondary peaks:} The portions of the widest separated CCF masked as containing the CCF's primary and secondary peaks are passed to the IDL routine MPFITFUN, which fits each peak separately with a Lorentzian model ($L = \rho\frac{\gamma^2}{(x-x_0)^2+\gamma^2}$, with $x_0$ the peak center, $\rho$ the peak height, and $\gamma$ the width parameter). From this fit we obtain initial values for the width, height, and location of each component's peak profile.

\item \textbf{Dual component fit to widest separated epoch:} Using the initial values determined for the peak parameters in the previous step, the widest separated CCF is refit using MPFITFUN with a dual Lorentzian model ($L_{CCF} = L_{prim} + L_{sec}$). The peak profile parameters (width, height, and peak center location) derived from this dual, simultaneous fit to a single CCF are used to place constraints on the width, relative peak heights, and maximum peak separation for all epochs for that source. We choose to constrain the width and relative peak height to $\pm 10\%$ of the best-fit values from the individual fits to the widest separated epoch. This step ensures that the peak parameters are fit consistently across all epochs for an individual source.

\item \textbf{Dual component fits to all epochs: } The full suite of CCFs measured for a given source are then fit with dual peak models. Each fit is initialized with the best-fit parameters from the dual component fit of the previous step, and variations in peak height and width are limited to $\pm 10\%$ of the best-fit values for the widest separated epoch: the primary degree of freedom at this stage is the peaks' locations, which are allowed to vary between epochs. The dual Lorentzian model is again fit with MPFITFUN, along with the aforementioned constraints. An example of a resulting fit is shown in Figure~\ref{fig:fits}.
\end{enumerate}

We applied this procedure to all $\sim 4500$ sources in the Complete IN-SYNC sample, producing radial velocity measurements for each component of the putative binary, along with statistics describing the structure of each CCF (i.e., the $r$ statistic, introduced in Equation \ref{eq:r}).

\subsection{Identifying candidate binaries} \label{selection}

To efficiently and objectively identify likely SB2s in the Complete IN-SYNC sample we identify parameters derived from the sources CCFs which can be used to pre-select likely SB2s for visual confirmation. For this purpose, we use three parameters -- two intrinsic and one model-dependent -- that can be calculated at each epoch for each source, and then used to eliminate sources with no evidence for a second spectroscopic component.

The two intrinsic properties of each source's CCF that we examine are $H$, the height of the primary CCF peak, and $r$, a parameter originally introduced by \citet{TonryDavis1979} to describe the strength of a given CCF peak relative to the CCF's anti-symmetric component. $H$ is measured as the maximum value of the CCF; low $H$ value indicates that the observed spectrum has low signal-to-noise ratio, or is poorly matched by the APOGEE template spectrum. The second intrinsic property that we measure, $r$, serves as a convenient tool to identify CCFs which are highly asymmetric due to the presence of a secondary peak. Following Eq. 23 by \citet{TonryDavis1979}, the $r$ parameter can be calculated as:
\begin{equation}\label{eq:r}
r = \frac{H}{\sqrt{2} \sigma_a}.
\end{equation}

\noindent where $\sigma_{a}$ gives the RMS of the anti-symmetric component of the CCF:

\begin{equation}\label{eq:sigma}
\sigma_a = \sqrt{\frac{1}{N} \sum \Big(c(n+x_0) - c(x_0 - n)\Big)^2},
\end{equation}

\noindent where $N$ is the number of lag steps in the total range, and $c(x_0 \pm n)$ signify the CCF function associated with the two halves of the CCF, divided about $x_0$, the location of the primary peak. Likely binaries with multiple CCF peaks will have low values of $r$ (high values of $\sigma_a$ due to a strong asymmetry from the secondary peak) and putatively single stars will have high values of $r$, due to their single, highly symmetric CCF peaks. To accentuate the asymmetry of the CCF further we calculate $r/\sigma_a$ for each epoch and retain the minimum value for each source. This minimum value corresponds to the epoch at which the CCF has the least symmetry about the primary peak. 

The final parameter we use to identify candidate SB2s is the velocity separation between the primary and secondary peaks identified in the automated process described in Section \ref{extraction}. Specifically, we identify candidate SB2s based on their maximum $v_{sep}$ measured across all epochs. Bona fide SB2s will have moderate maximum velocity separations, neither too large nor too small: too large a $v_{sep,max}$ indicates that the secondary radial velocity has been measured from a spurious/noisy CCF peak at a non-physical RV, and too small a $v_{sep,max}$ indicates that the `secondary' CCF peak is not separable from the primary peak itself.

To validate the range of $H$, $r$, and $v_{sep}$ values that can be used to pre-select a sample of likely SB2s, we visually inspected CCFs for hundreds of APOGEE targets to establish validated samples of high confidence SB2s, and sources with no evidence of binarity.  The first classification was performed on a sample of 200 sources: 50 APOGEE targets previously identified by S.D.C. as likely SB2s, to demonstrate the $H$, $r$, and $v_{sep}$ values of high confidence binaries, and 150 sources drawn randomly from the Complete IN-SYNC sample, to identify the parameter values of SB2s that can be identified with typical IN-SYNC spectra. The CCFs of sources in this semi-random sample were scrutinized carefully to identify less obvious SB2s: sources with subtle, but time-dependent CCF asymmetries that indicate the presence of two components.

Based on the properties of the SB2s we identified in the semi-random sample, we defined criteria which enable the pre-selection of a sample of potential SB2 targets.  Specifically, we preserve as viable SB2 candidates those sources that meet the following criteria:

\begin{itemize}
\item{$H > 0.30$ across all epochs; sources with low CCF peaks are discarded, as this indicates either a low S/N spectrum or a poor match between the APOGEE template spectrum and observed spectrum, either of which will pose challenges for reliable SB2 identification.}
\item{$(r/\sigma_a)_{min} < 460$; sources that do not meet this criteria are highly symmetric at all epochs, consistent with being a single, RV-stable star.} 
\item{$30 < v_{sep, max} < 220$ km s$^{-1}$; sources with maximum velocity separations above and below this threshold have non-physical putative secondary components, indicating the absence of a physically meaningful secondary component. }
\end{itemize}

Applying these criteria to the Complete IN-SYNC sample identified a set of $533$ candidate binaries ($\sim 12 \%$ of the Complete IN-SYNC sample).  

To evaluate the success and efficiency of these criteria for pre-selecting SB2s, two of us (M.A.F. and K.R.C.) then visually examined and classified all $533$ candidate SB2s, along with 10 additional candidates identified by a cursory examination of the CCFs for the Complete IN-SYNC sample. The CCFs of these $543$ candidate SB2s were examined and assigned a numerical value to rank the strength of the evidence for their binary status, which could include: at least one visit with fully separated peaks, multiple epochs with merged structures, or a majority of epochs with consistent shoulder behavior (see Figure~\ref{fig:fits}). Sources with fully separated peaks were rated the highest, and are almost certainly binary systems, while sources with either of the other two behaviors were rated lower and categorized as potential binaries. Based on this criteria scores from $1-5$ were assigned to candidate SB2s, where $1$ indicated low likelihood of binarity and $5$ indicated high likelihood of binarity. The two independent scores were then added to give each candidate SB2 a relative likelihood of being a bona fide SB2 on a $10$ point scale.

\begin{figure*}
\centering
\includegraphics[width = 0.8\hsize]{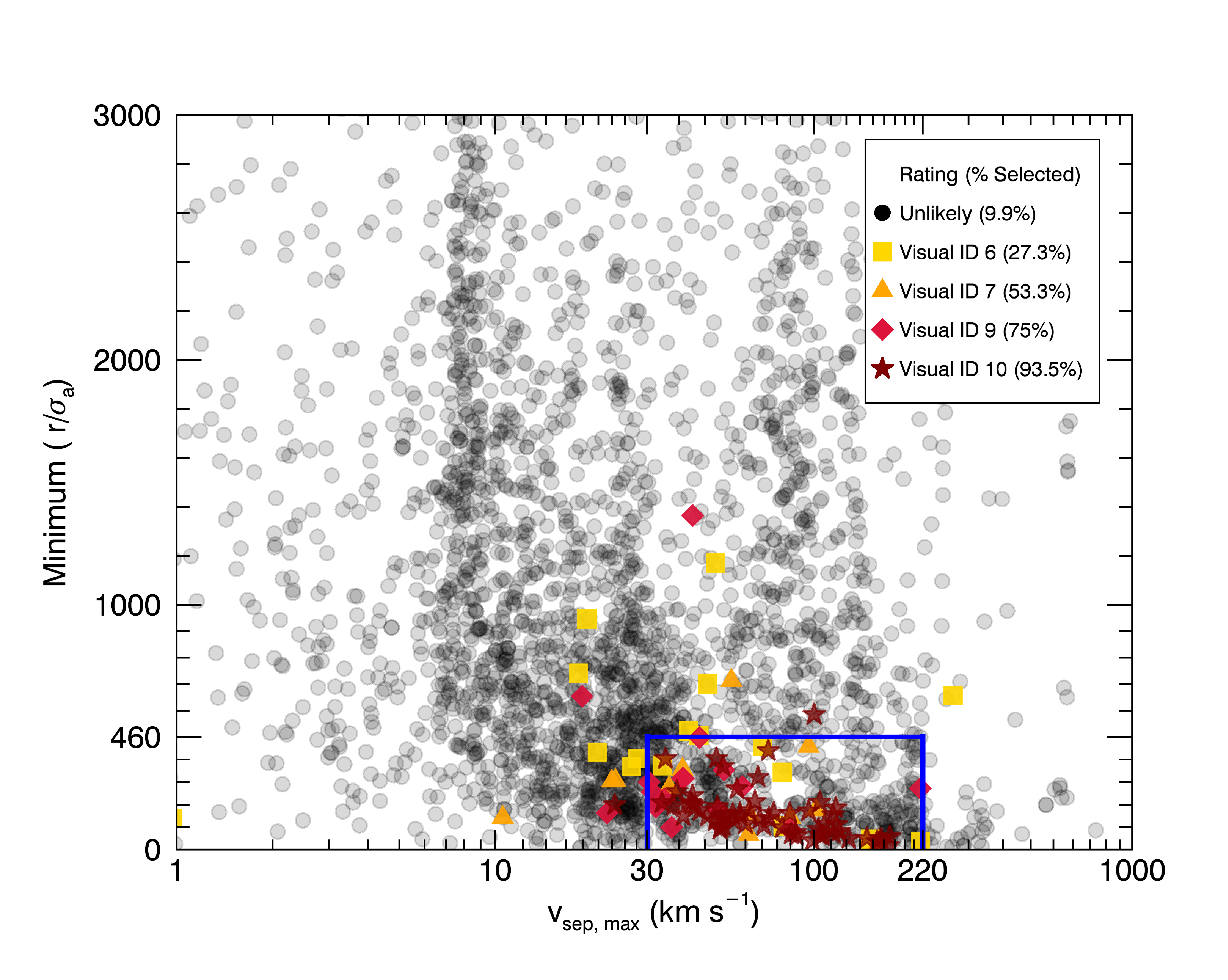}
\caption{Each complete IN-SYNC sample source's minimum $r/\sigma_a$ plotted as a function of its maximum velocity separation, v$_{sep,max}$. Ratings are described in Section \ref{selection} and run from 10 (high confidence) to 6 (potential). None of the visually rated sources had a cumulative rating of 8. Parenthetical percentages indicate the number of visually confirmed binary systems that satisfy the parameter cuts described in Section \ref{selection}: $30 <$ v$_{sep,max} <$ 220 km s$^{-1}$, $(r/\sigma_a)_{min} <$ 460 and H $> 0.30$ (not shown). Visual identification ratings lower than 6 were not included in the final sample.}
\label{fig:selection}
\end{figure*}

Figure \ref{fig:selection} shows the $543$ candidate SB2s selected from the Complete IN-SYNC sample, where each source has been color-coded to indicate its visual inspection rating.  We retain sources with a combined numerical rating of six or larger as binary candidates: we identify 70 high confidence binaries and 34 potential binaries. We identify systems with at least one epoch with fully separated peaks as high confidence binaries, and label as potential binaries those that show consistent merged structures or peak shoulder behavior throughout all epochs. Examples of these three behaviors can be seen in Figure \ref{fig:fits}. The percentage of visually classified sources retained by the parameter cuts outlined above are listed in the legend of Figure \ref{fig:selection}; these cuts retain almost all of the visually confirmed high confidence binaries, and most of the potential binaries.

It should be noted that the sample returned after the parameter cuts has a majority of false-positives ($439/533$), however the purpose of the parameter cuts is to reduce the number of sources which are visually inspected without losing likely SB2s. In this respect the parameter cuts do well, returning $> 90\%$ of the binaries which can be found by visual inspection of the entire sample, while reducing the number of stars that actually require that visual inspection by $\sim 90\%$. There are two primary reasons for these false-positives; low signal-to-noise epochs leading to poorly fit CCFs, and secondary peaks fit to noise ``spikes'' in the absence of a true second peak. The former case often gives poorly fit RVs for both the primary and the secondary component, while the latter often gives non-physical secondary RVs.

Sources selected as potential or high confidence binaries will often have epochs where the primary and secondary components have identical RVs, and thus the CCF has only one genuine peak: in this case, our peak fitting algorithm will fit a spurious secondary peak. In these cases, and when the object has multiple other well fit epochs, we adopt the primary RV for the secondary component as well. In future work the processes of Sections \ref{extraction} \& \ref{selection} will be swapped, such that candidate binaries will be identified based on direct quantitative measures of CCF structure, rather than fit parameters. This is due to the aforementioned issues of low signal-to-noise epochs, spurious secondary fits, and very broad peaked CCFs, all of which can lead to false indicators of binarity in the fit parameters.

\subsection{Velocity Residuals} \label{residuals}

\begin{figure}
\centering
\includegraphics[width = 0.8\hsize]{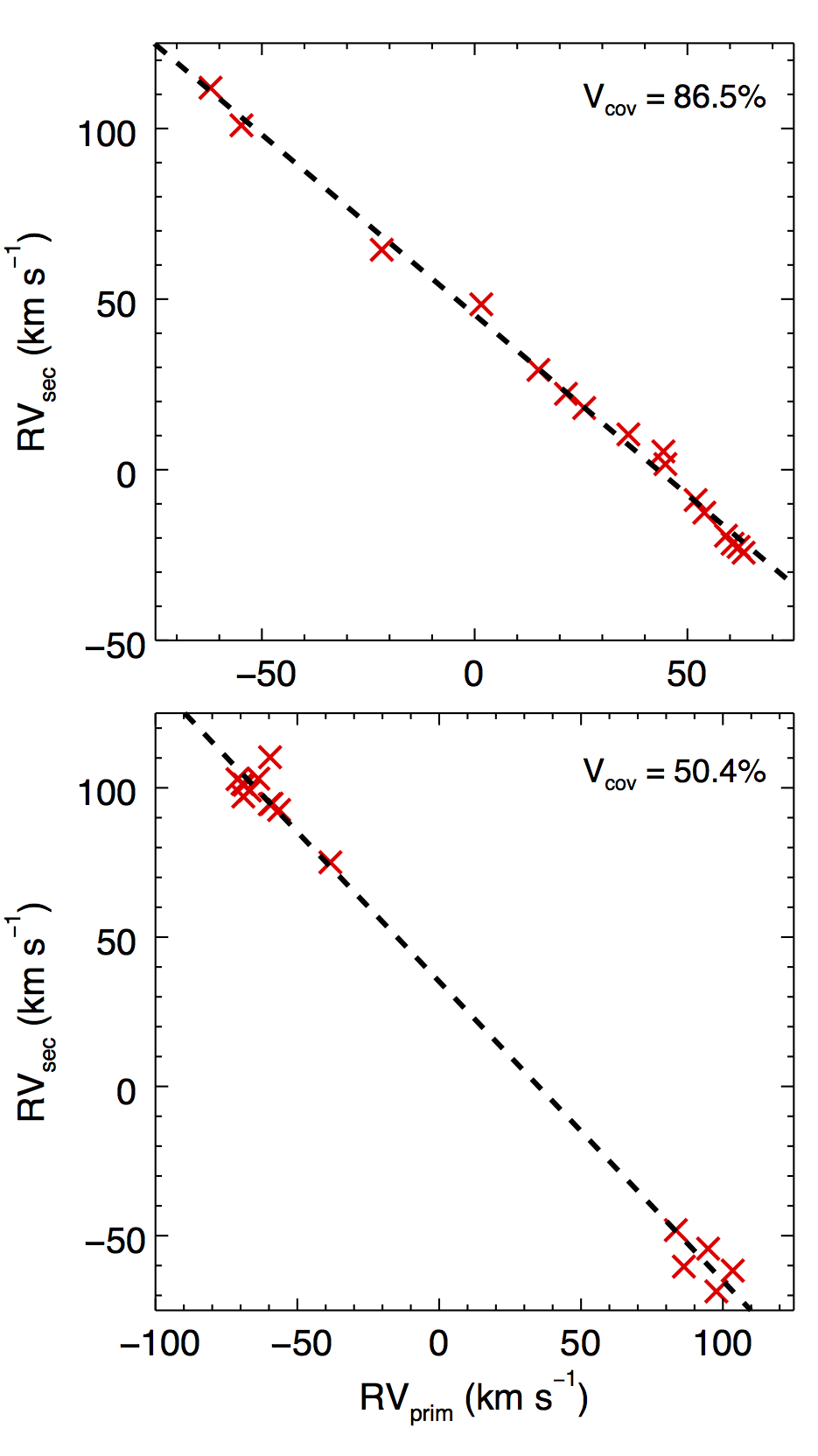}
\caption{RV$_{sec}$ as a function of RV$_{prim}$ for two of the most heavily visited sources in our sample. Linear fits to this data allows us to infer mass ratios and systemic velocities for a given system, and to characterize uncertainties in the RVs we extract for SB2s in our sample. The top panel is an example of a system which has adequate orbital coverage and is likely to have a converging orbital fit. The bottom panel is an example of poor orbital coverage, despite a high number of epochs. Orbital parameters fit to the bottom system are degenerate, with no clear best fit.}
\label{fig:linfits}
\end{figure}

\begin{figure}
\centering
\includegraphics[width = \hsize]{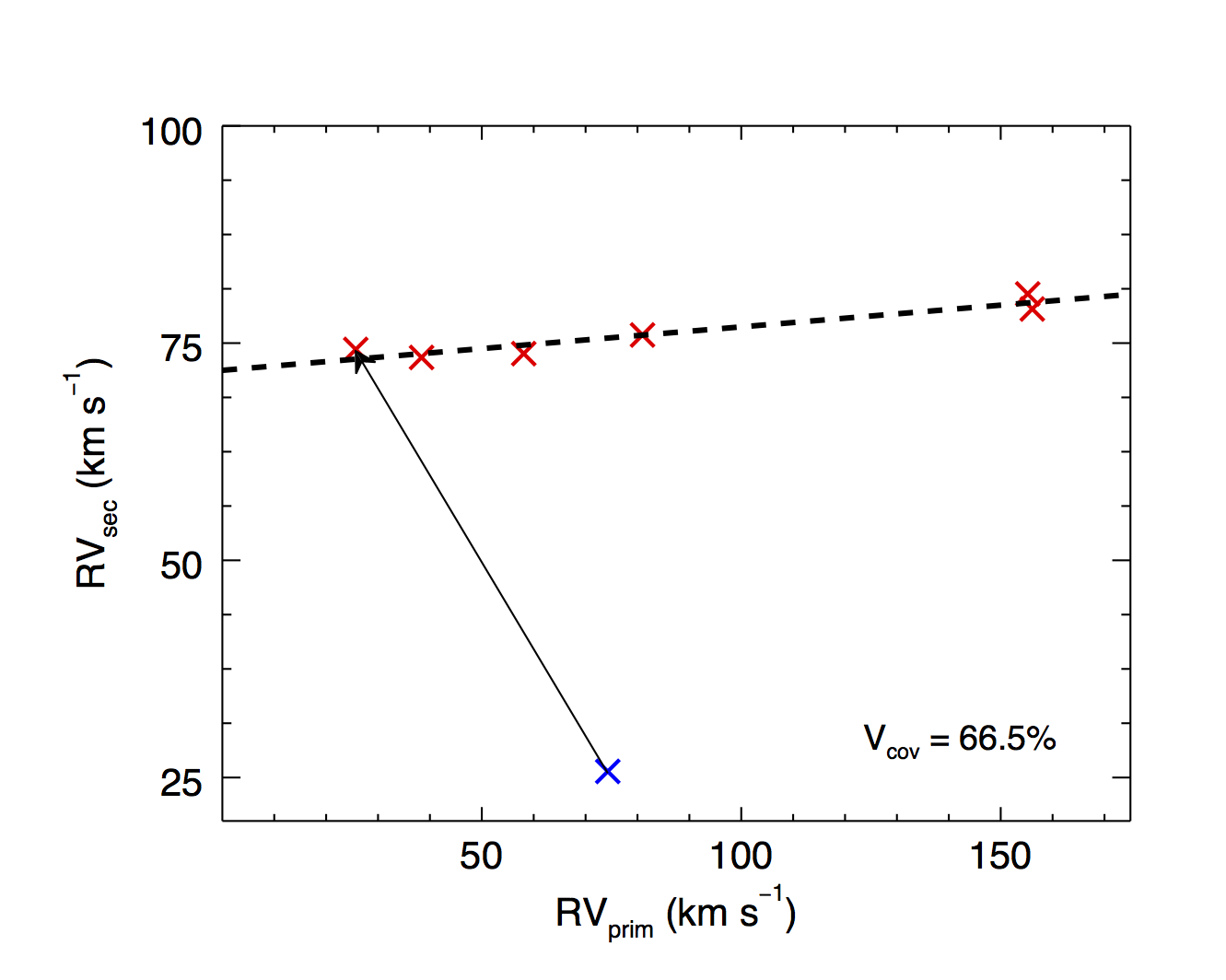}
\caption{Wilson plot for the high mass ratio system, 2M06413207+1001049. The primary component is identified based on the height of the CCF peak, not on the RV amplitude. The quality of the extracted RVs, along with the clear trend in this figure and the agreement for low mass ratios between our distribution and the SB9 sample indicate that this is a bona fide low mass ratio system. The blue data point is an example of our primary and secondary RVs being incorrectly assigned, and the arrow indicates where the data point moves when the assignments are flipped.}
\label{fig:highq}
\end{figure}

The Lorentzians fit to each CCF diagnose the RVs of each component, but do not provide a robust uncertainty estimate for those RVs. For single stars, centroiding techniques can routinely achieve velocity precisions of a few percent, but RVs inferred from multi-component fits such as we adopt here are subject to systematic and random errors that produce non-Gaussian error distributions that are best bounded with empirical error estimates. As an empirical estimate of the precision of our dual-component RV measurements, we report the residuals of our observed component velocities with respect to the best fit $RV_{prim}$ vs. $RV_{sec}$ (Wilson plot) for each system; examples of Wilson plots are shown in Figure \ref{fig:linfits} and Figure \ref{fig:highq}. Using a simple $\chi^2$ minimization method, we obtain the best linear fit to the $RV_{prim}$ vs $RV_{sec}$ data for each system. For each pair of RV measurements extracted from a given visit spectrum, we calculate the perpendicular distance between the location defined by those empirical measurements and the linear fit to the full ensemble of RV measurements of that system. We use this `perpendicular velocity distance' as a pseudo-measure of the uncertainty in our extracted radial velocities.

\begin{figure}
\centering
\includegraphics[width = \hsize]{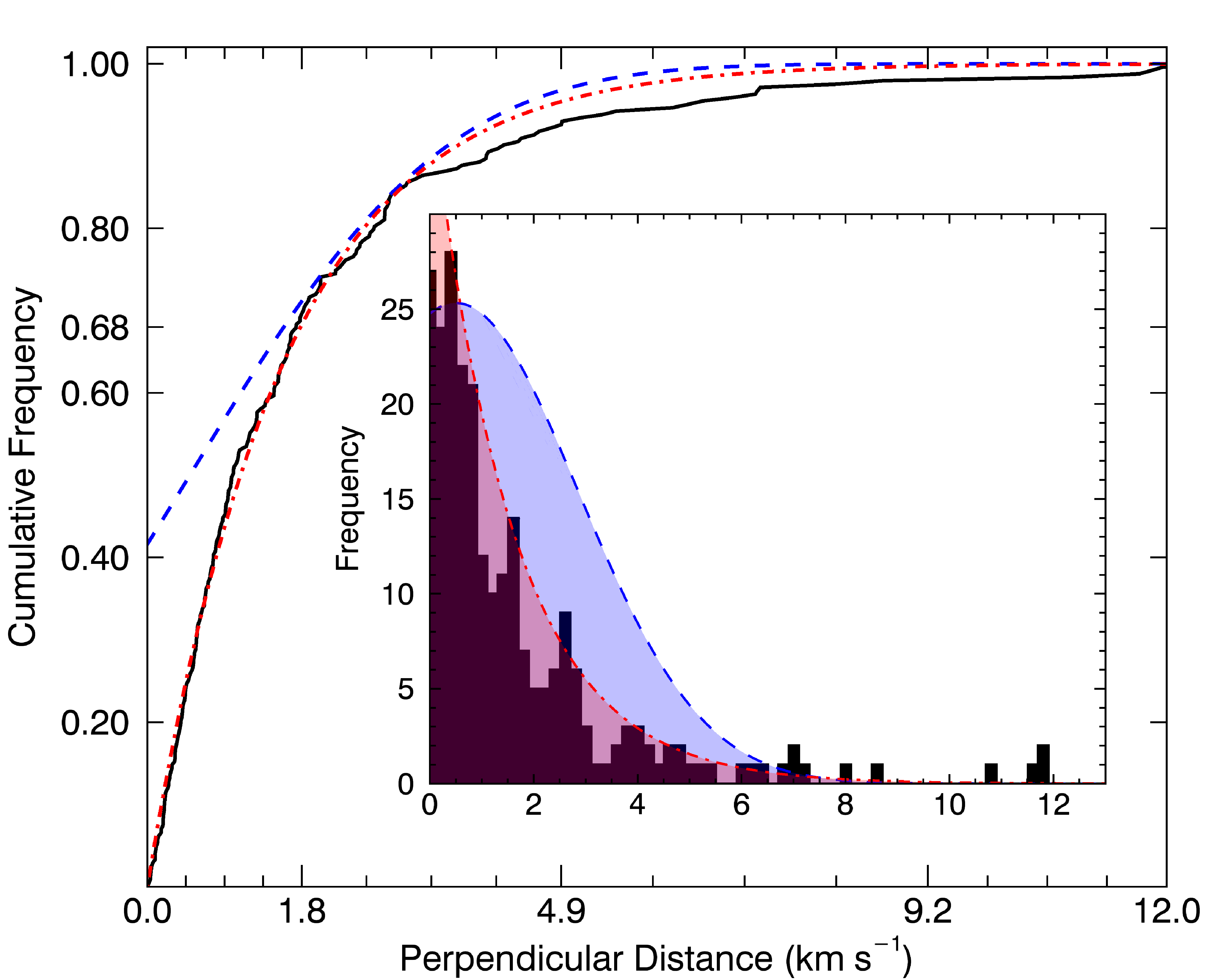}
\caption{Cumulative distribution of all perpendicular velocity distances for the systems with three or more epochs (black histogram), with best fit normal and exponential distributions (blue and red distributions, respectively). The perpendicular distance is detailed in Section \ref{residuals}. The exponential distribution is chosen as the best fit and its $1\sigma$ value, $1.817$ $km$ $s^{-1}$, is adopted as the residual for systems with $\le 2$ epochs. Inset: Histogram of the distribution of 3+ visit perpendicular velocity distances with fits overlaid.}
\label{fig:error}
\end{figure}

For systems with three or more visits, we calculate the `perpendicular velocity distance' for each epoch, and use that value as an estimate of that epoch's velocity precision. For systems with fewer than three epochs, we estimate the global precision of our RV measurements using a distribution fit to the velocity residuals calculated for $3+$ epoch systems. Figure \ref{fig:error} shows the distribution of the residuals for systems with three or more epochs, with normal and exponential distributions overlaid. The exponential fit better reproduces the empirical distribution than the normal fit across the range of velocity residuals. We adopt the exponential fit value corresponding to the $68$th percentile or pseudo-normal $1\sigma$ (1.817 km s$^{-1}$) as representative of the typical velocity errors for systems in our sample with fewer than three visits. Using this exponential fit to calculate the expected $95$th and $99$th percentile velocities deviations produces pseudo-normal errors of $2\sigma = 4.871$ and $3\sigma = 9.190$ km s$^{-1}$, respectively.

\section{Results} \label{results}

We present the radial velocities of 104 binary systems (70 high confidence, 34 potential), along with their velocity residuals, which serve as a measure of uncertainty for these velocity measurements. A literature search identifies 18 of these systems as previously known spectroscopic binaries; notes on previous identification of these systems, along with alternate IDs are in the Appendix. The radial velocities measured for each SB2 are given in Table \ref{table:rvs}, with 2MASS identifiers, Julian dates, and residuals for all epochs of all 104 candidate binary systems.

\begin{deluxetable}{lcccc}
\tablewidth{0pt}
\tabletypesize{\scriptsize}
\tablecaption{Radial Velocities}
\tablehead{
\colhead{2Mass ID} &
\colhead{Epoch (JD)} &
\colhead{RV$_p$} &
\colhead{RV$_s$} &
\colhead{$\delta^a$}
}
\startdata
03434101+3237320 &  2456168.941     &     14.987     &     29.258    &     0.405\\
---------------  &  2456170.899     &    -62.069     &    111.934    &     0.443\\
---------------  &  2456171.976$^f$ &    -21.820     &     64.460    &     0.984\\
---------------  &  2456172.974     &     21.460     &     22.245    &     0.448\\
---------------  &  2456174.926     &     59.062     &    -19.346    &     1.280\\
---------------  &  2456235.774$^f$ &      1.558     &     48.476    &     2.866\\
---------------  &  2456262.683     &     51.944     &     -8.891    &     0.638\\
---------------  &  2456282.621     &     62.035     &    -22.612    &     1.330\\
---------------  &  2456283.624     &     53.968     &    -12.576    &     0.393\\
---------------  &  2456561.992     &    -54.796     &    100.929    &     2.622\\
---------------  &  2456669.674$^f$ &     36.124     &     10.362    &     0.890\\
---------------  &  2456673.564     &     60.611     &    -21.697    &     1.748\\
---------------  &  2456675.565$^f$ &     25.740     &     18.096    &     0.528\\
---------------  &  2456701.589     &     44.359     &      5.358    &     2.841\\
---------------  &  2456703.590     &     63.206     &    -24.313    &     1.632\\
---------------  &  2456706.602     &     44.824     &      1.648    &     2.610\\
03444495+3213364 &  2456171.976     &     2.3097     &     29.077    &     0.816\\
---------------  &  2456235.774     &     16.011     &     16.244    &     0.875\\
---------------  &  2456675.565     &     15.375     &     14.849    &     0.528\\
---------------  &  2456703.590$^f$ &     -3.012     &     36.981    &     0.469
\enddata
\tablecomments{A sample of the RV Table available in full, online. Radial velocities are reported in pairs, in km s$^{-1}$, with perpendicular velocity distances also in km s$^{-1}$.}
\tablenotetext{a}{$\delta$ is the  perpendicular velocity distance from the best fit line in the RV$_{p}$ vs. RV$_{s}$ plot, as discussed in Section \ref{residuals}.}
\tablenotetext{f}{This flag denotes epochs where the RV assignment was flipped.}
\label{table:rvs}
\end{deluxetable}

Requiring the detection of spectroscopic features from both the primary and secondary biases this search, and all searches for SB2 systems, towards systems with comparable luminosities in the observed bandpass. Observations at infrared wavelengths reduce this bias relative to searches at optical wavelengths, by minimizing the luminosity difference due to the primary/secondary temperature contrast, but cannot eliminate the effect entirely. As a result, the sample of SB2s identified here includes a large number of systems with equivalent infrared luminosities, and thus comparably sized peaks within the system's CCF. As the relative heights of the CCF peaks provide our primary means of distinguishing the system's primary and secondary components, unambiguously assigning the RVs measured at a given epoch to a system's primary and secondary components is difficult in cases where the two components have CCF peaks of comparable height. The assignments can be improved, however, by examining the relationship between the primary and secondary RVs across all epochs: in many cases incorrect RV assignments are clearly visible as outliers from the best linear fits described in Section \ref{residuals} and seen in Figure \ref{fig:highq}. We have visually examined the relationship between each system's primary and secondary velocities, and adjusted assignments where doing so would improve the consistency of the relationship between the primary and secondary velocities. We include a flag in Table \ref{table:rvs} to identify all epochs where we have manually re-assigned the primary and secondary RVs in this manner. 

Using the RV measurements we have extracted for these systems, we estimate mass ratios, systemic velocities and, where possible, fit orbits to each system. To check cluster membership we compare the derived systemic velocities to their presumed clusters radial velocity. Inferred mass ratios are described in Section \ref{massratios}, while systemic velocities ($\gamma$) and cluster membership are described in Section \ref{systemic} and Section \ref{cluster}, respectively. Section \ref{orbits} details the method used to fit orbits to two systems from the final sample.

\subsection{Mass Ratios} \label{massratios}

Though absolute masses cannot be determined for non-eclipsing systems from spectroscopy alone, the system's mass ratio can be derived. We infer mass ratios for systems with more than a single epoch from the best linear fit to the system's RV$_{prim}$ vs RV$_{sec}$ relationship. Mass ratio estimates are the slope of the linear best fit to the RVs,

\begin{equation}
    \frac{M_{sec}}{M_{prim}} = \frac{RV_{prim}}{RV_{sec}}.
\end{equation}

\begin{figure*}
\centering
\includegraphics[width = 0.8\hsize]{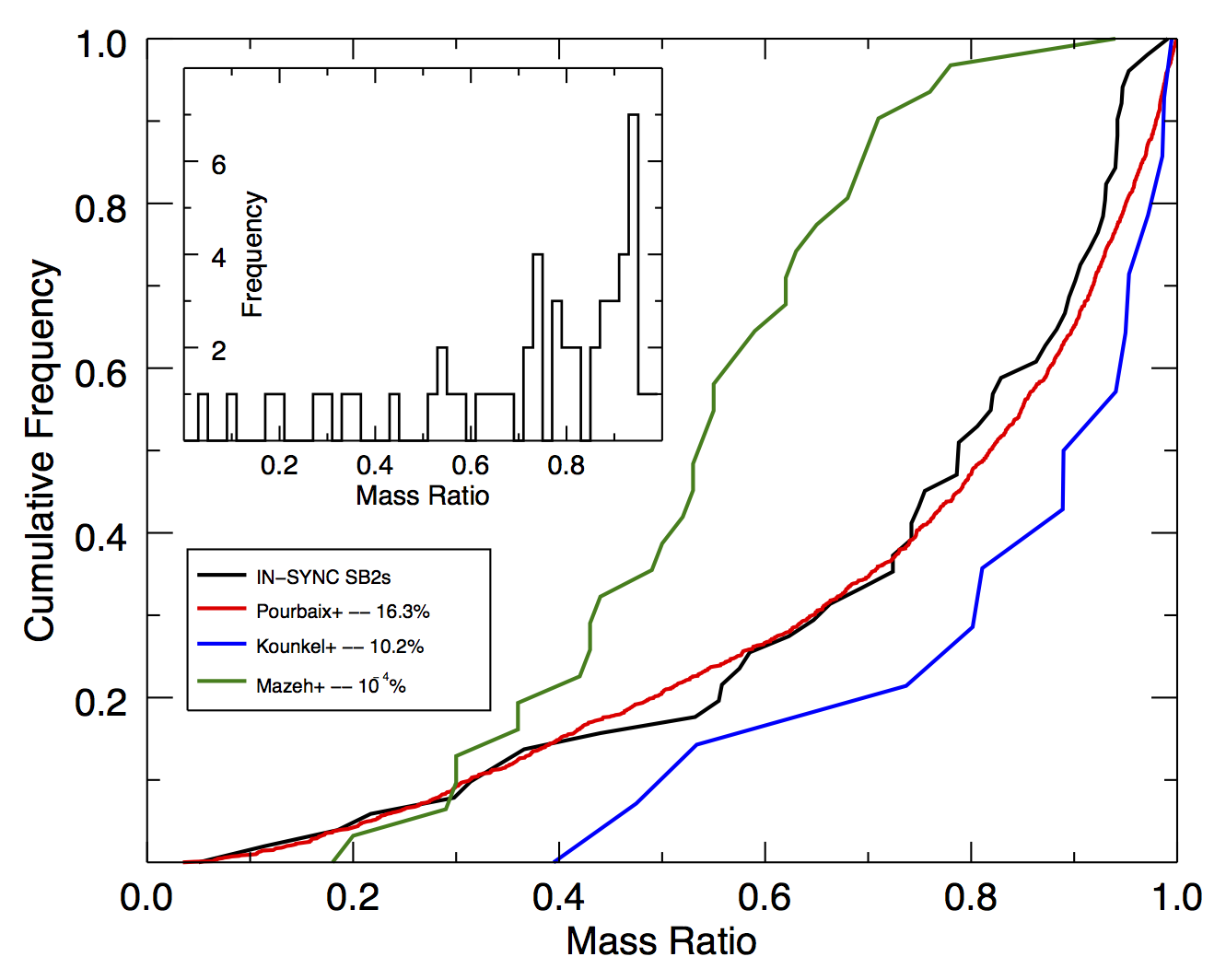}
\caption{CDFs for the IN-SYNC SB2 mass ratios along with the mass ratios from \citet{Pourbaix2004}, \citet{Mazeh2003}, and \citet{Kounkel2016}. Values given in the legend are the K-S statistic, or the probability that the IN-SYNC sample comes from the same mass ratio distribution as the Pourbaix+, Kounkel+, and Mazeh+ samples. The inset shows the mass ratio distribution for all $52$ systems in our sample with two or more epochs. As expected the majority of the systems are clustered near one.}
\label{fig:qcdfs}
\end{figure*}

The distribution of inferred mass ratios is shown in the inset of Figure \ref{fig:qcdfs} for systems with more than one epoch. As expected, the distribution is dominated by sources with mass ratios near one, but there are a number of sources with mass ratios $<$ 0.7. A number of these apparently low mass ratio systems have only a few observations ($2-3$ epochs), such that their inferred mass ratios may be particularly sensitive to errors in the extracted RV values, and/or the association of those RVs with the system's primary and secondary components. Nonetheless, there are several apparently low mass ratio systems with observations over several epochs and reliable RV$_{prim}$ vs. RV$_{sec}$ relationships, suggesting that they are bona fide low mass ratio systems (see Figure \ref{fig:highq}): three particularly notable examples are 2M03424086$+$3213347 ($16$ epochs and $q = 0.217$), 2M06413207$+$1001049 ($6$ epochs and $q = 0.050$), and 2M03430679$+$3148204 ($15$ epochs and $q = 0.340$).

We used a Kolmogorov-Smirnov test to determine if these low mass ratio systems were consistent with three prior measurements of the distribution of mass ratios in samples of SB2 systems\footnote{Note that \citet{Kounkel2016} and \citet{Pourbaix2004} did not report mass ratios for the systems in their catalogs.  We calculated mass ratios for these systems using the primary and secondary velocities, and velocity amplitudes, respectively, that those authors report.}. We compare the mass ratios we measure to those measured for: 1) $1410$ SB2s in the SB9 catalog of spectroscopic binaries \citep{Pourbaix2004}, 2) $32$ main-sequence SB2s characterized by \citet{Mazeh2003}, who used infrared spectroscopy to detect secondaries for optically identified SB1s, and 3) $15$ SB2s identified in the optical by \citet{Kounkel2016}. Figure \ref{fig:qcdfs} shows the CDFs for these three mass ratio distributions, along with the CDF for the mass ratios that we measure from the APOGEE/IN-SYNC sample.

We find a vanishingly small likelihood (K-S statistic$\sim 10^{-4}$\%) that the mass ratios of the IN-SYNC and Mazeh+ SB2s are drawn from the same distribution. This may reflect a bias against equal-mass systems in the \citet{Mazeh2003} sample: since $q=1$ systems would be unlikely to appear as an optical SB1, the sample of stars \citet{Mazeh2003} were able to convert into infrared SB2s should have preferentially lower $q$ values than an unbiased sample of SB2s. We find better agreement (K-S statistic$\sim 10$\%) between the IN-SYNC SB2 and Kounkel+ samples, with the Kounkel+ sample shifted towards higher mass ratio systems than the IN-SYNC SB2 sample. This is consistent with the difference in observations; Kounkel+ used optical spectroscopy, where low-mass secondaries will be difficult to detect, whereas this study used infrared spectroscopy, for which the contrast ratio between a hot primary and a cool secondary is more favorable. Surprisingly, we find our best agreement (K-S statistic$\sim 16$\%) between the IN-SYNC SB2 and Pourbaix+ samples. We also note the agreement between our sample and the SB9 sample is especially good for q $< 0.4$, indicating that the number of low mass ratio systems in our sample is consistent with this larger sample.

\subsection{Systemic Velocities} \label{systemic}
A binary's systemic velocity provides an important constraint on its potential membership in a cluster population.  We infer systemic velocities for all systems in our sample with multi-epoch coverage by using the best-fit RV$_{prim}$ vs. RV$_{sec}$ relationship to identify the point at which RV$_{prim} =$ RV$_{sec}$. As with the mass ratios determined from the same data, we include systemic velocities for binaries with two or three epochs but note that they are significantly more uncertain due to the sparse coverage of the RV$_{prim}$ vs. RV$_{sec}$ plane. The uncertainties we report for these systemic velocity estimates are propagated from the formal uncertainties in the slope and intercept of the best fit line.

Table \ref{table:params} contains the basic properties we have inferred for each system in our sample, including mass ratio and systemic velocity estimates. The table also includes the number of APOGEE spectra obtained for the system, as the reliability of the inferred parameters scale strongly with the temporal coverage. We remind the reader that mass ratios and systemic velocities are impossible to derive for sources with only a single epoch of APOGEE observations, and are highly uncertain for sources with only two or three epochs of observations. The confidence flag encodes the robustness of the binary identification, as outlined in Section \ref{selection}.

\begin{deluxetable}{lcccc}
\tablewidth{0pt}
\tabletypesize{\scriptsize}
\tablecaption{Fit Parameters}
\tablehead{
\colhead{2Mass ID} &
\colhead{N$_{epochs}$} &
\colhead{$\gamma$ (km s$^{-1}$)} &
\colhead{$q$ $\big(\frac{M_1}{M_2}\big)$} &
\colhead{C$^{^a}$}
}
\startdata
03434101+3237320 & 16 &    22.107 $\pm$   1.601 &    0.942$_{-0.018}^{+0.018}$ & L\\
03424086+3213347 & 16 &    22.743 $\pm$   3.143 &    0.217$_{-0.023}^{+0.023}$ & P\\
03450783+3102335 &  7 &   -62.974 $\pm$  11.080 &    0.788$_{-0.089}^{+0.089}$ & L\\
05342386-0515403 &  5 &    34.418 $\pm$   3.170 &    0.724$_{-0.024}^{+0.024}$ & L\\
03443444+3206250 &  2 &    23.622 $\pm$   6.450 &    0.819$_{-0.188}^{+0.181}$ & P\\
03292204+3124153 &  1 &      \nodata            &           \nodata            & L\\
\enddata
\tablecomments{A sample of the fit parameter table, which is provided in full online. Values for $\gamma$ and $q$ are not provided for single-visit systems, as they can only be determined from multi-epoch sources.}
\tablenotetext{a}{Confidence; $L$ indicates likely binary systems, $P$ potential binary systems.}
\label{table:params}
\end{deluxetable}

\subsubsection{Cluster Membership} \label{cluster}
\begin{figure}
\centering
\includegraphics[width = 0.9\hsize]{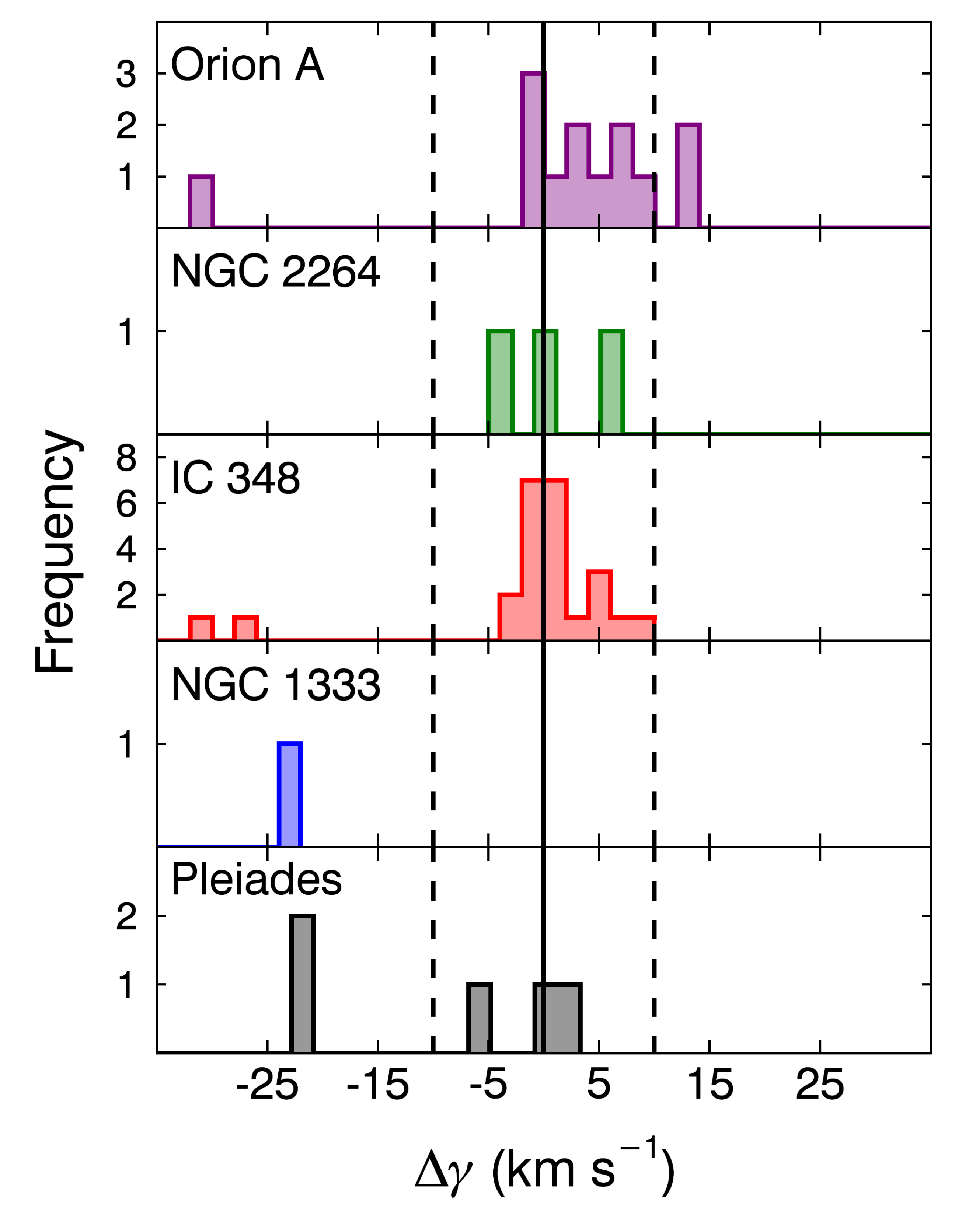}
\caption{Relative systemic velocities ($\Delta \gamma = \gamma - RV_{cluster}$) for systems in the final sample. The heliocentric radial velocities (in km s$^{-1}$) adopted for each cluster are; $RV_{Orion} = 25$ \citep{DaRio2016}, $RV_{2264} = 22$ \citep{Furesz2006}, $RV_{348} = 16$ \citep{Cottaar2015}, $RV_{1333} = 16$ \citep{Foster2015}, and $RV_{Pleiades} = 6$ \citep{Mermilliod2009}. The dashed lines show an RV range of $\pm 10$ km s$^{-1}$ around the cluster's central velocity, comparable to an $\sim$3 $\sigma$ error for a velocity measurement in our sample. Note that not all systemic velocities are shown here as several lie well away from their associated cluster's RV.}
\label{fig:systemics}
\end{figure}

Binaries that are physically bound to a cluster must have systemic velocities consistent with the cluster's mean velocity and overall velocity dispersion; if not, they would kinematically separate in short order. To evaluate membership status, we compare the systemic velocity measured for each binary in our sample to the radial velocity of the cluster to which it is presumed to belong. Figure \ref{fig:systemics} shows the velocity offset between the systemic velocities measured for binaries in each IN-SYNC cluster field and that cluster's mean RV.  We conservatively adopt a threshold of $\pm$ 10 km s$^{-1}$, corresponding to a $\sim$ 3$\sigma$ error for a typical individual velocity measurement in our sample (Section \ref{residuals}), to identify systems with systemic velocities consistent with membership in each cluster. Of the $52$ systems with more than one epoch of APOGEE spectra, we find 38 have systemic velocities that appear consistent with membership. Specifically, we find $22/26$ binaries in the IC 348 field have a systemic velocity consistent with membership; similarly, we find $10/14$, $3/6$, $3/4$, and $0/2$ binaries in the Orion A, Pleiades, NGC 2264 and NGC 1333 fields are likely members of the targeted stellar population. We also calculate the velocity dispersion of these likely cluster members and compare with values reported in the literature for each cluster: $\sigma_{348} = 3.23$ km s$^{-1}$ \citep[$0.72$ from][]{Cottaar2015}, $\sigma_{Orion} = 3.46$ km s$^{-1}$ \citep[$2.5$ from][]{DaRio2017}, $\sigma_{2264} = 6.03$ km s$^{-1}$ \citep[$3.5$ from][]{Furesz2006}, and $\sigma_{Pleiades} = 3.53$ km s$^{-1}$ \citep[$1.02$ from][]{Mermilliod2009}. In all cases, the dispersions we measure from the SB2 population exceed those measured previously for the not-obviously-RV variable cluster members. The large velocity dispersions are almost certainly a product of the larger uncertainties associated with the systemic velocities we measure, however; the uncertainties on the systemic velocities are typically at the few-km s$^{-1}$ level, of the same order as our measured velocity dispersions and significantly larger than the uncertainties associated with the non-variable sources in each cluster.

In the case of Orion A we can compare our simple kinematic membership determination with the membership probabilities calculated by \citet{Bouy2014} from astrometric and photometric measurements. Of the $65$ binaries we identify in the Orion A fields, $14$ have counterparts within $3$ arcseconds in the Bouy et al. catalog (note that this is not the same $14$ systems as those in the previous paragraph). Only three of these $14$, however, have multiple APOGEE observations, as required to estimate their systemic velocities.  Of those three, all have systemic velocities that meet our conservative criteria for membership status, although two have relatively large $\delta$RV values: 2MJ05345563$-$0601036 ($\gamma = 26.01 \pm 5.06$, Bouy probability $12.8\%$), 2MJ05352989$-$0512103 ($\gamma = 32.46 \pm 4.33$, Bouy probability $1.8\%$), 2MJ05364717$-$0522500 ($\gamma = 33.42 \pm 7.92$, Bouy probability $7.1\%$).

The remaining $11$ systems are single epoch sources, so we can do no better than to check if the component RVs bound the mean RV of Orion. In this regard $10/11$ of the single epoch systems are potential Orion members, though \citet{Bouy2014} infer a $0\%$ membership probability for $9/10$ of these systems. For the $11$th system, 2MJ05341347$-$0423539, we find component RVs of $38.2 \pm 1.8$ km s$^{-1}$, and $56.5 \pm 1.8$ km s$^{-1}$, which are both higher than the mean RV of the Orion population, while \citet{Bouy2014} infer a membership probability of $42.7\%$. Probabilities for all $14$ systems can be found in the Appendix.

\subsubsection{IC 348 Velocity Dispersion}\label{ic348dispersion}

For the 22 binaries with systemic velocities consistent with membership in IC 348, we can test if the dispersion in our derived systemic velocities is consistent with the value of 0.72 km s$^{-1}$ measured by \citet{Cottaar2015} from the full APOGEE dataset for this cluster. To do so, we compare the dispersion measured from our binary sample with that predicted by simulations incorporating the \citet{Cottaar2015} dispersion and the uncertainties in our measured systemic velocities. This simulation has three steps:
\begin{enumerate}
    \item Initialize a sample of 22 synthetic systemic velocities, drawn randomly from a normal distribution with a deviation equal to the dispersion reported in \cite{Cottaar2015}.
    \item To each synthetic systemic velocity, we add a randomly generated velocity error; each source's errors are sampled separately from normal distributions with standard deviations scaled to match the uncertainties associated with our individual systemic velocity measurements.
    \item We then compute the standard deviation of the ``measured'' systemic velocities in this synthetic cluster sample.
\end{enumerate}

\begin{figure}
\centering
\includegraphics[width = \hsize]{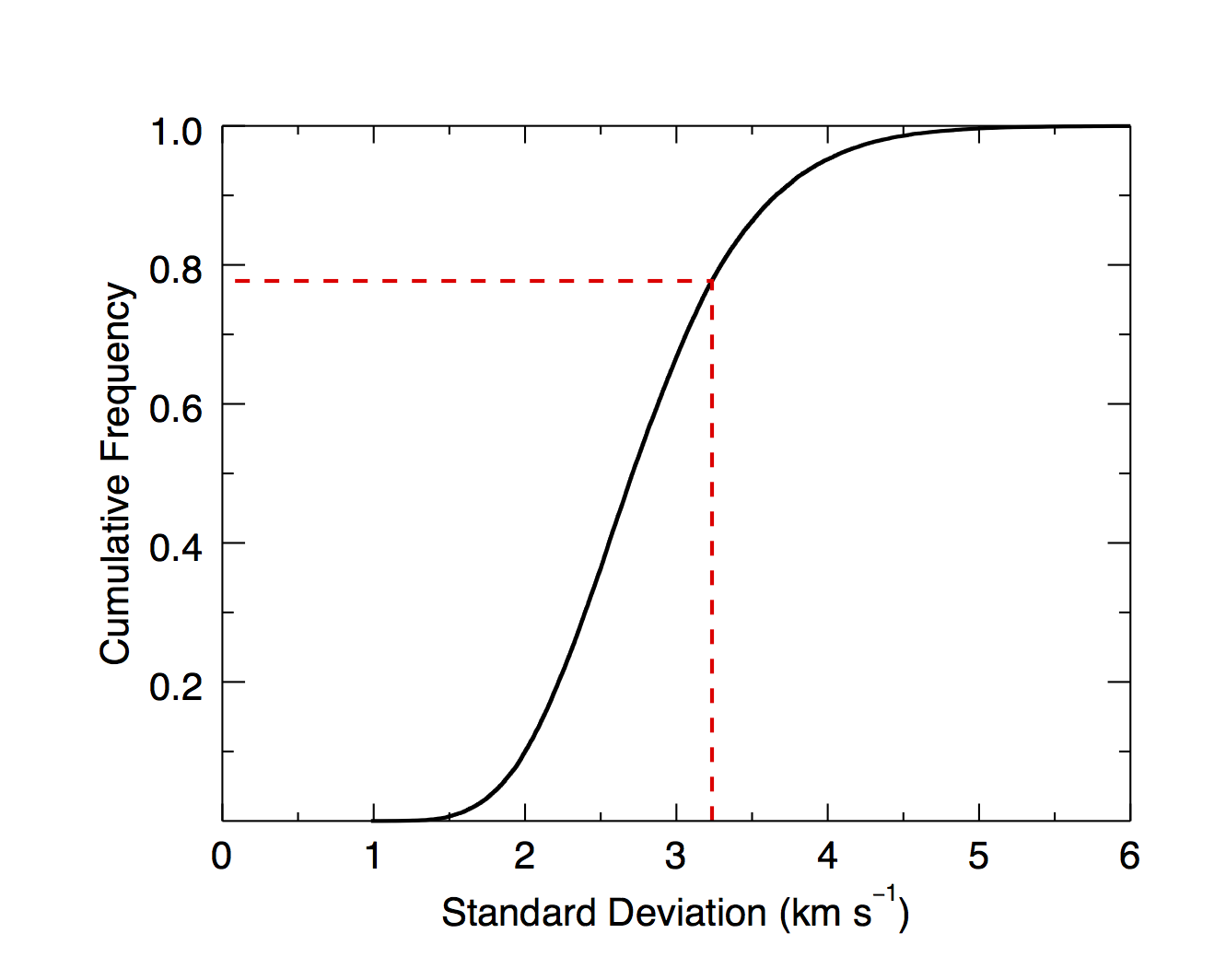}
\caption{Cumulative distribution of the systemic velocity dispersions simulated for 50,000 synthetic binary populations. Simulated populations are created such that the systemic velocities are consistent with the dispersion measured by \citet{Cottaar2015} for IC 348, with simulated uncertainties scaled to match our individual systemic velocity measurements. Our measured systemic velocity standard deviation is $3.23$ km s$^{-1}$, which falls at the $77.8$th percentile of the simulated distribution (shown in red).}
\label{fig:dispersion}
\end{figure}

Given the moderate size of our observed sample (22 binaries) we repeat this simulation 50,000 times to determine the distribution of dispersions we could expect to measure from our sample. We find a mean ``measured'' dispersion of $2.79$ km s$^{-1}$, with a standard deviation of $0.66$ km s$^{-1}$. The actual systemic velocities we measure for IC 348 members have a standard deviation of $3.23$ km s$^{-1}$, which is well within one standard deviation of the mean of the simulated set. Figure \ref{fig:dispersion} shows the cumulative distribution of the synthetic velocity dispersions measured from all 50,000 simulated systemic velocity samples. Our measured dispersion falls at the $77.8$th percentile, such that we could expect to measure a larger dispersion $\sim 22\%$ of the time. We conclude that the systemic velocities we measure are consistent with the dispersion measured by \citet{Cottaar2015}, but note that the width of the velocity dispersion measured from our binary sample is still dominated by our measurement errors; more accurate measurements are required to rule out potential differences in the velocity dispersions of single \& multiple cluster members.

\subsection{Orbit Fits} \label{orbits}
We attempt to fit orbits to systems with at least eight epochs of APOGEE spectra that meet our criteria for RV coverage and period significance. Our criteria for acceptable RV coverage is based on a statistic developed by \citet{Troup2016} to describe the fraction of a star's velocity range which is well sampled by the available RV data:
\begin{equation}
    V_{cov} = \frac{N}{N-1}\left(1 - \frac{1}{RV^2_{span}}\sum_{i=1}^N     \left(RV_{i+1}-RV_i\right)^2\right),\label{eq:velcov}
\end{equation}

\noindent where $N$ is the number of epochs, $RV_{span} = RV_{max}-RV_{min}$, and $RV_i$ are the extracted radial velocities (Equation 23, \citet{Troup2016}). We calculate this statistic using the primary and secondary RVs measured for each system and use the average of these two values to characterize the quality of the system's velocity coverage. The utility of the V$_{cov}$ velocity coverage parameter can be seen in Figure \ref{fig:highq} (V$_{cov} = 66.5$\%) and Figure \ref{fig:linfits} (top panel: V$_{cov} = 86.5$\%; bottom panel: V$_{cov} = 50.4$\%). We only attempt to fit orbits to systems with V$_{cov} > 80\%$; systems with less coverage are prone to erroneous fits to an eccentric orbit with a prominent velocity maxima near periastron in the unsampled portion of the orbit.

We also limit our orbit fits to sources whose period determinations have false alarm probabilities $<$3\%. We measure periods by computing a Lomb-Scargle periodogram over a range of periods tuned to the temporal sampling of each system: the periodogram is computed for periods as long as four times the time between the minimum RV and the nearest RV measurement which differs by $80$\% of the total RV span ($RV_{max} - RV_{min}$). We use this initial periodogram to identify the most likely/significant periods, and recompute their significance with more densely sampled periodograms computed for a narrow period range ($\delta P \pm 2\%$), and retain the period with the lowest false alarm probability.

Taken together, our criteria for performing a full orbital fit are: APOGEE RVs from at least eight epochs (of which there are ten sources), velocity coverage (V$_{cov}$) $> 80$\%, and a period with a false alarm probability of $3$\% or less. Only two systems from our final sample meet all three requirements for a secure orbital fit: 2M03423215+3229291 ($N$ = 16, velocity coverage $87.1$\%, period significance $97.9$\%) and 2M03434101+3237320 (16, $86.5$\%, $98.2$\%).

Two other systems meet our velocity coverage criteria, and have periods with significance $>$ $90$\%: 2M03441568+3231282 and 2M03430679+3148204. The first of these suffers from extremely similar primary and secondary peak heights in the CCFs, making the assignment of RVs ambiguous at all epochs. The second system only exhibits shoulder behavior, where the primary and putative secondary CCF peaks are merged at all epochs, such that our RV extraction method fails to provide robust primary and secondary RVs.
    
\begin{figure}
\centering
  \begin{tabular}{@{}cc@{}}
    \includegraphics[width=0.95\hsize]{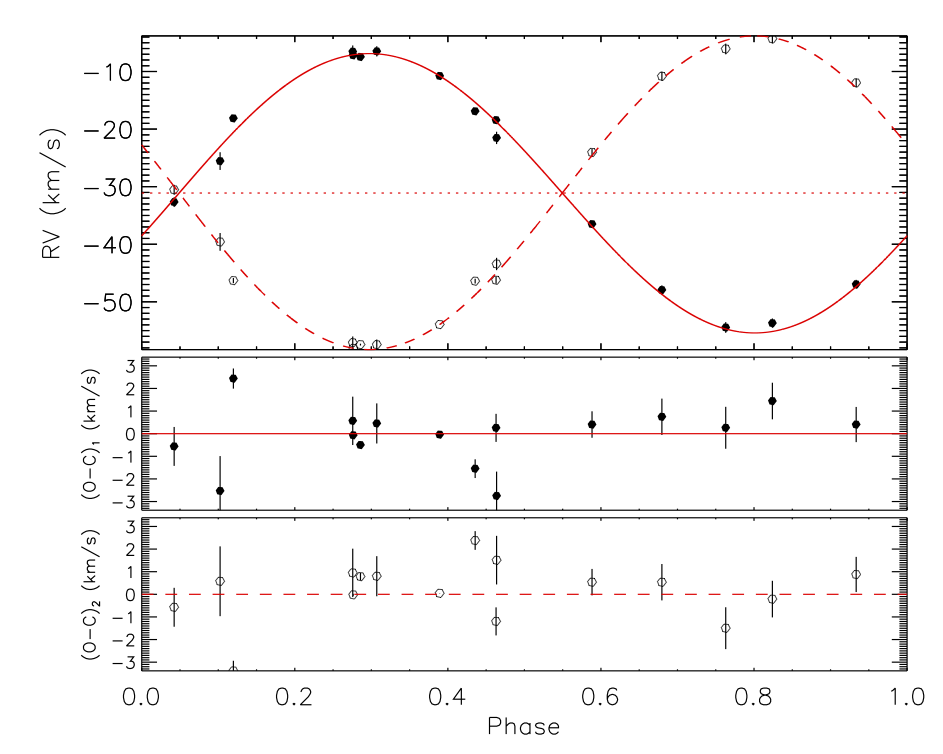} \\
    \includegraphics[width=0.95\hsize]{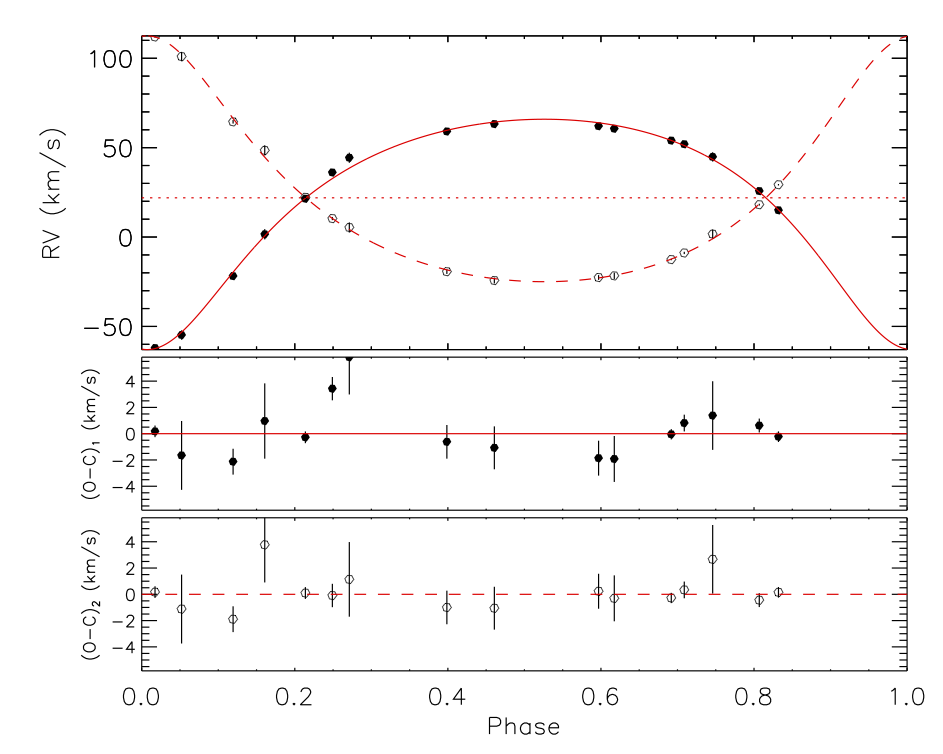}
  \end{tabular}
  \caption{Two systems with orbit fits using RVfit from \citet{rvfit}. Top panel: 2MJ03423215$+$3229291. Bottom panel: 2MJ03434101$+$3237320.}
  \label{fig:orbits}
\end{figure}

We fit orbital parameters to these two systems using the IDL routine RVfit by \citet{rvfit}. RVfit uses a simulated annealing algorithm, which is a Monte-Carlo method relying on random sampling, to fit Keplerian orbits to sets of RVs. The orbital parameters that we determine for these two systems, along with their associated uncertainties, are given in Table \ref{table:orbits}. Phased orbits are plotted in Figure \ref{fig:orbits}.

\begin{deluxetable}{lcccc}
\tablewidth{0pt}
\tabletypesize{\scriptsize}
\tablecaption{Orbital Parameters}
\tablehead{
\colhead{2Mass ID} &
\colhead{03423215+3229291} &
\colhead{03434101+3237320}
}
\startdata
Period (days)         & 2.372 $\pm$ 0.001   & 10.560 $\pm$ 0.001 \\
T$_p$ (HJD)           & 2456169.5 $\pm$ 0.1 & 2456170.7 $\pm$ 0.1\\
Eccentricity          & 0.006 $\pm$ 0.005   & 0.319 $\pm$ 0.001  \\
$\omega$ (deg)        & 252.29 $\pm$ 1.36   & 174.91 $\pm$ 0.71  \\
K$_{1}$ (km s$^{-1}$) & 24.26 $\pm$ 0.08    & 64.42 $\pm$ 0.28   \\
K$_{2}$ (km s$^{-1}$) & 27.25 $\pm$ 0.07    & 68.74 $\pm$ 0.28   \\
q (M$_2$/M$_1$)       & 0.890 $\pm$ 0.004   & 0.937 $\pm$ 0.006  \\
$\Delta$q             & 0.007               & 0.005              \\
$\gamma$ (km s$^{-1}$)&-31.10 $\pm$ 0.07    & 21.93 $\pm$ 0.07   \\
$\Delta \gamma$       & 0.05                & 0.177              \\
\enddata
\tablecomments{$\Delta$q and $\Delta \gamma$ are the difference between the mass ratio and systemic velocity derived using the Wilson plots and those returned from the RVfit routine. All four measures lie within the combined uncertainties.}
\label{table:orbits}
\end{deluxetable}

Note that RVfit requires the user to define upper and lower bounds for the seven fit parameters (systemic velocity, eccentricity, period, angle and time of periastron, and component velocity amplitudes, $K_1$ and $K_2$.) and additionally returns an estimate for the mass ratio. The upper and lower bounds for each orbital parameter were:
\begin{itemize}
    \item Eccentricity: zero to one.
    \item Angle of periastron: zero to $2\pi$ radians.
    \item Time of periastron: the full time span of the source's observations.
    \item Systemic velocity: $\gamma \pm \delta$, where $\gamma$ and $\delta$ are our systemic velocity estimate and uncertainty, respectively (Section \ref{systemic}).
    \item Velocity amplitudes: $RV_{max}+10$ km s$^{-1}$ and $RV_{min}-10$ km s$^{-1}$, where $RV_{min}$ and $RV_{max}$ are the minimum and maximum RVs measured, regardless of component designation.
    \item Orbital Period: 0 to 4 times the greater of the primary and secondary time spans computed as the time between the minimum RV and the RV which exceeds the minimum RV by at least $80$\% of the total RV span ($RV_{max} - RV_{min}$).
\end{itemize}

Both sources are located in the IC 348 field, however the systemic velocities obtained from the orbit fit (in good agreement with those obtained from the Wilson plot) indicate that neither system is a member of the IC 348 cluster. Both systems have mass ratios implying near equal mass components, with photometric colors suggesting an early M spectral type for 2M03423215+3229291 (J-K = 0.88), and spectra in the literature indicating a solar-like primary for 2M03434101+3237320 \citep[G0]{Nesterov1995}. Both systems also have relatively short periods, but differ in eccentricity; 2M03423215+3229291 is nearly circular whereas 2M03434101+3237320 is decidedly non-circular, with a robust eccentricity of e$\sim$0.3. We note that \citet{CiezaBaliber2006} measure a 2.3d photometric period for 2M0342315+3229291 which agrees quite well with our spectroscopically determined period, suggesting that the system may be either tidally distorted or eclipsing.

\section{Conclusions} \label{conclusion}

We have used cross-correlation functions, a data product of the APOGEE pipeline, to identify and extract component radial velocities for $104$ highly confident or potential double-lined spectroscopic binaries (SB2s) in the APOGEE/IN-SYNC fields. To our knowledge, 86 of these are newly identified SB2s. Sources for the 18 previously detected binary systems are given in the Appendix along with notes on individual sources of interest.

\begin{itemize}

\item Dual Lorentzians were fit to APOGEE CCFs to extract pairs of radial velocities for each epoch for all sources in the Complete IN-SYNC sample ($\sim 4500$ sources). Cuts applied to two CCF parameters and one radial velocity parameter were used to create a sub-sample of the likeliest binaries. This sub-sample was visually inspected to eliminate false positives and create a final sample of potential and high confidence binary systems. Uncertainty in the extracted radial velocities was estimated using linear fits to the radial velocity data, and for low epoch sources a statistical expectation value.

\item Using the extracted radial velocities, we inferred mass ratios and systemic velocities for systems with at least two epochs from the best linear fit to the RV$_{prim}$ vs. RV$_{sec}$ data. The distribution of mass ratios is skewed towards $q = 1$, likely a product of a detection bias towards high mass ratio systems with larger velocity separations. Systemic velocities measured for multi-epoch systems were used to evaluate the likelihood that each system is a member of the cluster population targeted in each field: the majority of the sample, particularly in the densest fields with higher temporal coverage, possess systemic velocities consistent with membership in the targeted stellar populations.

\item For two systems with at least eight epochs of RV measurements, good velocity coverage (V$_{cov} > 80\%$) and a robust period measurement (significance $>$ 97\%), we use the RVfit toolkit to fit the system's RVs and derive a full set of orbital parameters. Both systems have near equal mass components, relatively short periods, and systemic velocities indicative of non-membership with IC 348. The two systems differ in eccentricity, with 2M03423215+3229291 nearly circular and 2M03434101+3237320 distinctly non-circular.

\item Using RV membership as a proxy for pre-main sequence status, we have identified 38 likely pre-main sequence SB2s, a number which will likely increase once systems identified from single-visit spectra are followed-up and confirmed as RV members. This sample is comparable to the entire population of 44 nearby (d $\lesssim$ 500 pc) pre-main sequence SB2s identified in the literature by \citet{Schaefer2014}. Optical multi-fiber surveys in Taurus and Chameleon \citep{Nguyen2012} and Orion \citep{Furesz2006,Kounkel2016} have recently produced similarly rich yields: \citet{Kounkel2016} identify 130 candidate single- and double-lined spectroscopic binaries, with 5 SB2s in Orion A and 10 in NGC 2264, of similar scale to our sample of 104 SB2s, including 10 systems in Orion A and 3 in NGC 2264. With APOGEE observations of nearby star-forming regions ongoing, we believe the population of well-studied pre-main sequence spectroscopic binaries will continue to expand quickly.

\end{itemize}

Work is ongoing to improve the methods developed here to identify and characterize SB2s with APOGEE observations, and to apply them to a broader cross-section of the APOGEE dataset. Computing the bisector of the CCF, and combining that measure of the CCF's symmetry with the $r$ parameter introduced earlier, shows promise for improving the yield of automatically selected candidate SB2s. Tests using dedicated MCMC methods to determine the significance and uniqueness of the period determined for a system, and its full orbital solution, also point to the ability to improve the reliability of the orbital parameters inferred for the systems with the highest number of APOGEE observations. We plan to apply these methods to first identify and characterize SB2s with the most promise for fitting full orbital solutions: systems with a large number of APOGEE observations (to improve the orbital coverage), a high contrast between the strength of the primary and secondary peaks (to remove the primary/secondary ambiguity), and/or high value classes of targets (ie, pre-main sequence stars in newly observed APOGEE fields; low-mass M/L dwarf stars; etc.). 

\textbf{Authorship Statement:} M.A.F. advanced and implemented the RV extraction algorithm, developed the criteria used to select the sample of 104 SB2s, calculated mass ratios and systemic velocities, implemented orbit fits, and drafted the manuscript. K.R.C. co-led the IN-SYNC collaboration, conceived and oversaw the identification and analysis of the IN-SYNC SB2s, developed the initial algorithm for extracting multiple RVs from APOGEE CCFs, and helped draft and edit the manuscript. 

N.D.L. contributed to the development of algorithms and criteria to identify SB2 candidates, and to perform and validate orbital fits. 

S.D.C. visually identified and cataloged numerous SB2 candidates with APOGEE observations, assisted in validating the properties of the IN-SYNC SB2s, and with G.Z., vetted target lists, finalized plate designs, oversaw the production of aluminum plug plates and integrated the IN-SYNC fields into the APOGEE observing schedule.

J.C.T. and M.R.M. conceived the IN-SYNC program's scientific motivation and scope, led the initial ancillary science proposal, oversaw the project's progress, and contributed to the interpretation of the findings discussed herein. N.D.R designed the observing strategy and led target selection in the Orion A fields; J.B.F and M.C. participated in sample selection of the IN-SYNC targets in the Perseus fields. R.B., A.M.R., G.W.R., and J.S. assisted in the construction of the appendix; K.S. and N.T. provided feedback on the analysis and presentation of the manuscript.   

\clearpage

\acknowledgments

\section{Acknowledgments}
We thank Marina Kounkel, Adam Kraus and Lisa Prato for useful discussions that improved our analysis and interpretation of the mass ratio distribution in the IN-SYNC fields. K.R.C. and M.A.F acknowledge support provided by the NSF through grant AST-1449476, and from the Research Corporation via a Time Domain Astrophysics Scialog award.

Funding for SDSS-III has been provided by the Alfred P. Sloan Foundation, the Participating Institutions, the National Science Foundation, and the U.S. Department of Energy Office of Science. The SDSS-III web site is http://www.sdss3.org/.

SDSS-III is managed by the Astrophysical Research Consortium for the Participating Institutions of the SDSS-III Collaboration including the University of Arizona, the Brazilian Participation Group, Brookhaven National Laboratory, Carnegie Mellon University, University of Florida, the French Participation Group, the German Participation Group, Harvard University, the Instituto de Astrofisica de Canarias, the Michigan State/Notre Dame/JINA Participation Group, Johns Hopkins University, Lawrence Berkeley National Laboratory, Max Planck Institute for Astrophysics, Max Planck Institute for Extraterrestrial Physics, New Mexico State University, New York University, Ohio State University, Pennsylvania State University, University of Portsmouth, Princeton University, the Spanish Participation Group, University of Tokyo, University of Utah, Vanderbilt University, University of Virginia, University of Washington, and Yale University.

The Two Micron All Sky Survey was a joint project of the University of Massachusetts and the Infrared Processing and Analysis Center (California Institute of Technology). The University of Massachusetts was responsible for the overall management of the project, the observing facilities and the data acquisition. The Infrared Processing and Analysis Center was responsible for data processing, data distribution and data archiving. 

This research has made use of NASA's Astrophysics Data System Bibliographic Services, the SIMBAD database, operated at CDS, Strasbourg, France, and the VizieR catalogue access tool, CDS, Strasbourg, France \citep{Ochsenbein2000}.

\clearpage

\appendix \label{appendix}

\LongTables

\begin{deluxetable}{lcccc}
\tablewidth{0pt}
\tabletypesize{\small}
\tablecaption{Appendix}
\tablehead{
\colhead{2Mass ID} &
\colhead{Alternate ID} &
\colhead{Bouy Membership$^a$} &
\colhead{$\gamma$ (km s$^{-1}$)} &
\colhead{Known?}
}
\startdata
03400448+3118568  &   BD+30 561  &   \nodata  &   17.439 $\pm$ 1.582  &   No\\
03405779+3118059  &   V900 Per  &   \nodata  &   18.441 $\pm$ 0.582  &   No\\
03423215+3229291  &     &   \nodata  &   -31.053 $\pm$ 2.035  &   No\\
03424086+3213347  &     &   \nodata  &   22.743 $\pm$ 3.143  &   No\\
03430073+3304482  &     &   \nodata  &   -14.629 $\pm$ 4.808  &   No\\
03430679+3148204  &     &   \nodata  &   14.749 $\pm$  2.199  &   No\\
03431992+3202412  &   LRL 1840  &   \nodata  &   19.391 $\pm$ 3.154  &   No\\
03434101+3237320  &   HD 281153  &   \nodata  &   22.107 $\pm$ 1.601  &   No\\
03435812+3213568  &   LRL 323  &   \nodata  &   -10.077 $\pm$ 50.870  &   No\\
03441143+3219401  &   LRL 137  &   \nodata  &   17.565 $\pm$ 1.498  &   No\\
03441568+3231282  &     &   \nodata  &   18.192 $\pm$ 1.078  &   No\\
03441776+3204476  &   LRL 169  &   \nodata  &   22.239 $\pm$ 4.387  &   No\\
03443444+3206250  &   LRL 198  &   \nodata  &   23.622 $\pm$ 6.450  &   Yes, D99\\
03443482+3211180  &   LRL 151  &   \nodata  &   16.442 $\pm$ 1.204  &   No\\
03443979+3218041  &   LRL 76  &   \nodata  &   17.820 $\pm$ 2.303  &   No\\
03444173+3212022  &   LRL 133  &   \nodata  &   15.378 $\pm$ 0.603  &   No\\
03444495+3213364  &   LRL 112  &   \nodata  &   15.501 $\pm$ 2.404  &   Yes, D99\\
03444508+3214130  &   LRL 138  &   \nodata  &   16.142 $\pm$ 2.375  &   No\\
03444770+3219117  &   LRL 17  &   \nodata  &   16.275 $\pm$ 1.757  &   No\\
03445064+3219067  &   LRL 3  &   \nodata  &   14.607 $\pm$ 1.247  &   No\\
03445561+3209198  &   LRL 50  &   \nodata  &   15.594 $\pm$ 0.490  &   No\\
03450783+3102335  &   Tyco 2356  &   \nodata  &   -62.974 $\pm$ 11.08  &   No\\
03452214+3202040  &     &   \nodata  &   17.512 $\pm$ 2.171  &   No\\
03475033+3226225  &   HD 281224  &   \nodata  &   15.883 $\pm$ 3.748  &   No\\
03485329+3132297  &     &   \nodata  &   17.737 $\pm$ 0.993  &   No\\
03490216+3242086  &   HD 281219  &   \nodata  &   26.672 $\pm$ 5.574  &   No\\
03250916+3126099  &     &   \nodata  &   -42.954 $\pm$ 2.077  &   No\\
03261467+3209453  &   HD 278694  &   \nodata  &   -6.707 $\pm$ 1.146  &   No\\
03292204+3124153  &     &   \nodata  &   \nodata  &   No\\
06400851+0944134  &   Cl*NGC2264 LBM1968  &   \nodata  &   30.009 $\pm$ 2.025  &   No\\
06410360+0930290  &   Cl*NGC2264 FMS2-1022  &   \nodata  &   22.293 $\pm$ 2.996  &   No\\
06413207+1001049  &   Cl*NGC2264 SBL393  &   \nodata  &   75.650 $\pm$ 7.098  & Yes, K16\\
06413433+0925533  &   Cl*NGC2264 DS469  &   \nodata  &   18.128 $\pm$ 4.694  &   No\\
05321683-0548237  &   BD-05 1284  &   \nodata  &   29.288 $\pm$ 4.593  &   No\\
05324407-0529523  &     &   \#269093  0\%  &   \nodata  &   No\\
05331624-0613195  &     &   \nodata  &   \nodata  &   No\\
05335130-0451477  &     &   \nodata  &   \nodata  &   No\\
05340693-0439303  &   Parenago 1346  &   \nodata  &   \nodata  &   No\\
05341347-0423539  &     &   \#336639 42.7\%  &   \nodata  &   No\\
05342386-0515403  &   V*V1676 Ori  &   \nodata  &   34.418 $\pm$ 3.170  &   No\\
05342626-0646093  &     &   \nodata  &   \nodata  &   No\\
05342767-0537192  &   V*V1962 Ori  &   \#346113 0\%  &   \nodata  &   No\\
05343087-0425070  &   V*V1973 Ori  &   \#348436 0\%  &   \nodata  &   No\\
05343903-0455288  &     &   \#354365 0\%  &   \nodata  &   Yes, T09\\
05344556-0529209  &   V*V1445 Ori  &   \nodata  &   \nodata  &   No\\
05344823-0447401  &   V*V1703 Ori  &   \#360509  &   \nodata  &   No\\
05345220-0440117  &   V*SX Ori  &   \#363141 0\%  &   \nodata  &   Yes, T09\\
05345249-0449404  &     &   \nodata  &   40.014 $\pm$ 8.882  &   No\\
05345431-0454129  &     &   \nodata  &   \nodata  &   Yes, M12\\
05345563-0601036  &   V*V1710 Ori  &   \#365208 12.8\%  &   26.006 $\pm$ 5.059  &   Yes, T09\\
05350138-0615175  &     &   \nodata  &   \nodata  &   No\\
05350200-0520550  &   V*V2118 Ori  &   \nodata  &   \nodata  &   No\\
05350326-0449209  &   V*V1718 Ori  &   \nodata  &   29,122 $\pm$ 4.263  &   No\\
05350392-0529033$^b$  &   V*V1481 Ori  &   \nodata  &   24.893 $\pm$ 3.012  &   Yes, T09\\
05350463-0509557  &   Haro 4-342  &   \nodata  &   \nodata  &   Yes, T09\\
05351077-0651557  &   Parenago 1831  &   \nodata  &   162.462 $\pm$ 186.936  &   No\\
05351113-0536511$^b$  &   V*V2212 Ori  &   \nodata  &   \nodata  &   Yes, C08\\
05351121-0517209  &   V*V2214 Ori  &   \nodata  &   \nodata  &   No\\
05351214-0531388$^b$  &   COUP 468  &   \nodata  &   \nodata  &   Yes, I07\\
05351236-0543184  &   V*V486 Ori  &   \nodata  &   \nodata  &   No\\
05351352-0527286  &     &   \#374677 0\%  &   \nodata  &   No\\
05351554-0525140$^b$  &   V*V1501 Ori  &   \nodata  &   \nodata  &   Yes, M12 \\
05351561-0524030  &   COUP 726  &   \nodata  &   \nodata  &   Yes, K16\\
05351580-0646291  &     &   \nodata  &   25.819 $\pm$ 4.111  &   No\\
05351755-0453037  &     &   \#377053  &   \nodata  &   No\\
05351798-0604430  &     &   \nodata  &   \nodata  &   No\\
05352147-0557421  &   Haro 4-359 SS  &   \#378903 14.5\%  &   \nodata  &   No\\
05352349-0520016  &   COUP 1249  &   \nodata  &   \nodata  &   No\\
05352813-0523064  &   V 2471 Ori  &   \nodata  &   \nodata  &   No\\
05352860-0455036  &   V 1739 Ori  &   \nodata  &   31.397 $\pm$ 3.423  &   Yes, T09\\
05352989-0512103  &     &   \#383142 1.8\%  &   32.460 $\pm$ 4.331  &   Yes, T09\\
05352992-0644151  &     &   \nodata  &   \nodata  &   No\\
05353002-0434276  &     &   \nodata  &   \nodata  &   No\\
05353154-0540278  &   V1560 Ori  &   \nodata  &   \nodata  &   No\\
05353645-0421304  &     &   \nodata  &   \nodata  &   No\\
05354261-0526083  &     &   \nodata  &   \nodata  &   No\\
05354311-0424556  &   V*BO Ori  &   \nodata  &   \nodata  &   No\\
05354337-0622195  &   V*V810 Ori  &   \nodata  &   \nodata  &   No\\
05354463-0450098  &   V1580 Ori  &   \nodata  &   \nodata  &   No\\
05355408-0528327  &   H97b 10831  &   \#396048 0\%  &   \nodata  &   No\\
05360185-0517365  &   V578 Ori  &   \nodata  &   \nodata  &   Yes, K16\\
05361723-0617245  &   V2657 Ori  &   \#410850 0\% (0.06\arcsec)    &   \nodata  &   No\\
                  &              &   \#410851 100\% (0.07\arcsec)  &   \nodata  &   No\\
05363704-0504412  &   V657 Ori  &   \nodata  &   \nodata  &   No\\
05364717-0522500  &   V2706 Ori  &   \#432270 7.1\% (2.7\arcsec)  &   33.422 $\pm$ 7.921  &   No\\
                  &              &   \#432353 98.7\% (2.8\arcsec) &                       &     \\
05365078-0459334  &   HD 294267  &   \nodata  &   \nodata  &   Yes, P04\\
05370100-0735469  &     &   \#441887 0\%      &   \nodata  &   No\\
05371161-0723239  &     &   \nodata  &   \nodata  &   No\\
05380160-0722218  &     &   \nodata  &   \nodata  &   No\\
05382589-0659519  &   V*V994 Ori  &   \nodata  &   -5.934 $\pm$ 1.827  &   No\\
05402047-0911572  &     &   \nodata  &   \nodata  &   No\\
05405530-0813169  &     &   \nodata  &   \nodata  &   No\\
05411524-0752111  &     &   \nodata  &   \nodata  &   No\\
05412451-0856320  &     &   \nodata  &   26.492 $\pm$ 6.987  &   No\\
05413037-0912268  &     &   \nodata  &   \nodata  &   No\\
05413338-0759562  &     &   \nodata  &   \nodata  &   No\\
05420151-0756153  &     &   \nodata  &   \nodata  &   No\\
05422393-0809459  &     &   \nodata  &   \nodata  &   No\\
05432464-0813275  &     &   \nodata  &   38.781 $\pm$ 8.232  &   No\\
03433710+2338322  &   Cl*Melotte22 DH278  &   \nodata  &   0.724 $\pm$ 0.139  &   No\\
03441396+2532155  &   V* V515 Tau  &   \nodata  &   7.516 $\pm$ 0.315  &   No\\
03454440+2413132$^b$  &   CI Melotte 22 761  &   \nodata  &   5.781 $\pm$ 2.145  &  Yes, M92\\
03505403+2309119  &   BD+22 578  &   \nodata  &   -16.397 $\pm$ 3.611  &   No\\ 
03510274+2325420  &   HD 24088  &   \nodata  &   -15.376 $\pm$ 8.875  &   No\\
03512664+2504369  &   TYC 1804-1166-1  &   \nodata  &   -56.807 $\pm$ 4.003  &   No\\
\enddata
\tablecomments{(C08) \citet{Cargile2008}; (D99) \citet{Duchene1999}; (I07) \citet{Irwin2007}; (K16) \citet{Kounkel2016}; (M92) \citet{Mermilliod1992}; (M12) \citet{Morales2012}; (P04) \citet{Pourbaix2004}; (T09) \citet{Tobin2009}}
\tablenotetext{a}{Sources matched to within $3\arcsec$, unless otherwise noted.}
\tablenotetext{b}{Systems with entries in our Notes on Individual Sources of Interest.}
\end{deluxetable}

\section{Notes on Individual Sources of Interest}

\textbf{2MJ03454440+2413132} -- AKA HII 761 and EPIC 211078009, a confirmed Pleiades member and known spectroscopic binary, first reported by \citet{Mermilliod1992} as a single-lined 3.3 day system with a G2 primary and a photometrically estimated mass ratio of q $=$0.65. To the best of our knowledge, however, no spectroscopic detections of the secondary component have been reported, consistent with the low mass ratio we measure from two epochs of APOGEE spectra, q $=$ 0.298. Velocities from one epoch of APOGEE data are not included in our analysis due, ironically, to the high contrast ratio between the (well-separated) CCF peaks.  The light curve of the system was also collected by the K2 mission during its monitoring of the Pleiades cluster; the system's three day period is visible in some reductions of the system's light curve, along with low frequency variations that are likely responsible for \citet{Rebull2016} classifying it as a non-periodic source in their analysis of the K2 light curves.

\textbf{2MJ05350392--0529033} -- AKA V1481 Ori, identified as a spectroscopic binary by \citet{Tobin2009}, and fully solved by \citet{Messina2016} as an orbitally synchronized M3+M4 system.  The mass ratio we measure from 2 epochs of APOGEE spectroscopy, q $=$ 0.55, agrees well with the value of q $=$0.54 measured by \citet{Messina2016}. 

\textbf{2MJ05351113--0536511} -- AKA Parenago 1802, a known pre-main sequence eclipsing binary system (see \citet{Cargile2008} and \citet{Gomez2012}.) 

\textbf{2MJ05351214--0531388} --  AKA JW 380 and COUP 468, identified as a low-mass eclipsing binary system by \citet{Irwin2007}.

\textbf{2MJ05351554--0525140} --  AKA V1501 Ori, identified by \citet{Morales2012} as a confirmed spectroscopic binary, and a potential eclipsing system with a period $\lesssim$ 24 days.  \citet{Kounkel2016} do not identify the source as an RV variable; with only one epoch of APOGEE spectroscopy, we cannot measure the system's mass ratio or characterize its orbit, but we can confirm that the system exhibits a velocity separation of $\approx$100 km s$^{-1}$.

\end{document}